\documentclass[12pt, draftclsnofoot, onecolumn]{IEEEtran}

\ifCLASSINFOpdf

\else

\fi
\usepackage[cmex10]{amsmath}
\usepackage{amssymb}
\usepackage{cite}
\usepackage{graphicx}
\usepackage{array,color}
\usepackage{amsmath}
\usepackage{stfloats}
\usepackage{graphicx}
\usepackage{subfigure}
\usepackage{tabularx}
\usepackage{epsfig,epsf,color,balance,cite}
\usepackage{algorithmic}
\usepackage{algorithm}
\usepackage{bm}
\usepackage{textcomp}

\begin{document}

\title{Joint User Selection and Energy Minimization for Ultra-Dense Multi-channel  C-RAN with Incomplete CSI}

\author{Cunhua Pan,  Huiling Zhu, Nathan J. Gomes and Jiangzhou Wang, \emph{Fellow}, \emph{IEEE}
\thanks{This paper has been accepted by IEEE JSAC with special issue on Deployment Issues and Performance Challenges for 5G. This work is supported by  European Commission Horizon2020 project iCIRRUS under grant agreement No 644526. }
\thanks{C. Pan, H. Zhu, N. Gomes and J. Wang are with the School of Engineering and Digital Arts, University of Kent, Canterbury, Kent, CT2 7NZ, U.K. C.Pan is also with the Queen Mary University of London, London E1 4NS, U.K. (Email:\{C.Pan, H.Zhu, N.J.Gomes, J.Z.Wang\}@kent.ac.uk).}
}

\maketitle
\vspace{-1.5cm}
\begin{abstract}
This paper provides a unified framework to deal with the challenges arising in dense cloud radio access networks (C-RAN), which include huge power consumption, limited fronthaul capacity, heavy computational complexity, unavailability of full channel state information (CSI), etc. Specifically, we aim to jointly optimize the remote radio head (RRH) selection, user equipment (UE)-RRH associations and beam-vectors to minimize the total network power consumption (NPC) for dense multi-channel downlink C-RAN with incomplete CSI subject to per-RRH power constraints, each UE's total rate requirement, and fronthaul link capacity constraints. This optimization problem is NP-hard. In addition, due to the incomplete CSI, the exact expression of UEs' rate expression is intractable. We first conservatively replace UEs' rate expression with its lower-bound. Then,  based on the successive convex approximation (SCA) technique and the relationship between the data rate and the mean square error (MSE), we propose a single-layer iterative algorithm to solve the NPC minimization problem with convergence guarantee. In each iteration of the algorithm, the Lagrange dual decomposition method is used to derive the structure of the optimal beam-vectors, which facilitates the parallel computations at the Baseband unit (BBU) pool. Furthermore, a bisection UE selection algorithm is proposed to guarantee the feasibility of the problem. Simulation results show the benefits of the proposed algorithms and the fact that a limited amount of CSI is sufficient to achieve performance close to that obtained when perfect CSI is possessed.
\end{abstract}
\vspace{-1.0cm}

\IEEEpeerreviewmaketitle

\section{Introduction}

The fifth-generation (5G) wireless system is expected to offer a thousand times the throughput \cite{Andrews2014} of the current fourth-generation (4G) \cite{huilingtcom09,huilingtcom12,huilingzhuJSAC} and provide ubiquitous service access for a large number of  user equipments (UEs) in hot spots such as shopping malls, stadia, etc. To achieve this goal, heterogeneous and small cell network (HetSNet) is regarded as one of the most promising techniques by exploiting spatial degrees of freedom through deploying more and more access points (APs) \cite{xiaohuge2016}. However, since all APs reuse the same frequency, the interference among the APs is a limiting factor \cite{xiaohuge2016}, which should be carefully managed. Dense cloud radio access network (C-RAN) was proposed in \cite{mugenpeng2014} as one promising architecture to conquer this issue. In dense C-RAN, all the base-band processing is performed at the BBU pool through the recent development of cloud computing techniques \cite{Rost2014}, while the RRHs are only responsible for simple radio transmission or reception \cite{huilingjsac2011,jiangzhou2012}. Due to their simple functionality, RRHs can be densely deployed in the network with low hardware cost. Due to the centralized architecture of dense C-RAN, the multi-UE interference can be efficiently handled through joint signal processing techniques such as coordinated multi-point (CoMP), leading to significant performance gains. Although C-RAN has been introduced in 4G, it is usually deployed in a large geographical area by connecting macrocell base stations to the BBU pool through fronthaul links. This conventional C-RAN  incurs large delays on the fronthaul links due to long transmission distance between RRHs and BBU pool \cite{mobile2011c}, which will violate the stringent latency requirement in 5G \cite{Andrews2014}, i.e., a roundtrip latency within 1 ms. In contrast, dense C-RAN studied in this paper is aimed to cover hot spots with much smaller geographical area. Hence, delays can be significantly reduced.

However,  there are many technical and deployment issues associated with dense C-RAN. First, dense deployment of RRHs will require high power consumption if all RRHs are activated even when the network traffic load is low. In addition, if each RRH serves all UEs, significant power will be used on the fronthaul links. As a result, how to activate the RRHs and select the RRHs for serving each UE to minimize the total network power consumption (NPC) is a critical issue. Second, in a dense C-RAN there will be a need for a large number of fronthaul links, requiring them to be low cost.  There may also be a need to use millimeter wave (mmWave) technology for flexible and low cost deployment.  These cost considerations lead to the likelihood of a capacity constraint on the fronthaul. Third, in dense C-RAN, the BBU pool will support large number of RRHs and the number of optimization variables for beam-vectors will become very large, which will incur high computational complexity and will become unaffordable. Finally, the dense C-RAN requires more CSI  for the facilitation of CoMP transmission design, which will cause a heavy training overhead. The amount of training overhead will increase with the number of RRHs and UEs, and may counteract the cooperative gains provided by CoMP transmission \cite{Caire2010}. The most promising way to deal with this issue is to restrict the number of RRHs that each UE should measure CSI to. The remaining CSI values can be regarded as zeros, or only long term channel statistics of the remaining CSI, such as path loss and shadowing, are considered. How to design transmission strategies for this incomplete CSI case  becomes an imperative task.

Most of  current work only deals with parts of the above challenges. For example, \cite{Yuanming2014,Shixin2015,dai2016energy,PAN17} considered the joint RRH selection and beamforming design to minimize the total NPC subject to UEs' quality of service (QoS) targets and per-RRH power constraints. These papers ignored the capacity constraints on the fronthaul links and assumed that the fronthaul capacity is unlimited. To address the fronthaul capacity constraints issue,  \cite{Jian2013} investigated the problem of minimizing the number of data transfers on the aggregated fronthaul links with UEs' QoS constraints and power constraints on each RRH.  However, \cite{Jian2013}  did not explicitly impose the fronthaul capacity constraints in the optimization problem. Recently, several papers have addressed the case when the fronthaul capacity constraints are explicitly imposed \cite{Binbin2014,dwkNg2015,Liu2016}. The case when the optimization problem is infeasible was not considered. Then, some UEs can be removed to make the optimization problem feasible again. The UE admission control and total NPC minimization were jointly optimized in \cite{Ha2014}, where a single-stage optimization problem was formulated by introducing a weighting factor in the admission control part. Recently, \cite{Abdelnasser2016} extended the work in \cite{Ha2014} to multi-channel heterogeneous  C-RAN where the C-RAN  is overlaid by a macro-cell.  However, for the admission control designs considered in \cite{Ha2014} and \cite{Abdelnasser2016}, one has to carefully choose the weighting factor associated with the admission control part to ensure that the selected UEs can satisfy the QoS constraints, which is not easy.

However, the algorithms proposed in \cite{Yuanming2014,Shixin2015,dai2016energy,PAN17,Jian2013,Binbin2014,dwkNg2015,Liu2016,Ha2014,Abdelnasser2016} were based on the assumption of full CSI at the BBU pool, which is not practical as explained. Unfortunately, the algorithms designed for perfect CSI cannot be directly extended to the case of incomplete CSI. To the best of our knowledge, only a few papers have considered the incomplete CSI case \cite{Shi2014ICC,Kim2014,Lakshmana2016,Liu2016tvt}. \cite{Shi2014ICC} proposed a CSI reduction scheme named compressive CSI acquisition, that can obtain the instantaneous CSIs for a subset of channel links and the large scale fading gains of the others. Based on the incomplete CSI, \cite{Shi2014ICC} solved a transmit power minimization problem while guaranteeing UEs' QoS requirements by using a stochastic coordinated beamforming technique. However, the method needs to solve a high-dimension semi-definite programming (SDP) problem for each sample, and the number of samples increases with the size of the network, which incurs an unacceptable complexity for dense C-RAN. \cite{Kim2014} focused on the beamforming algorithm to maximize the sum-rate  for arbitrary UE-centric clustering C-RAN. The ``C-cluster method'' was introduced in \cite{Kim2014} to reduce channel estimation overhead where only subsets of CSIs for each UE are measured, and the other unavailable CSIs are regarded as zeros.  Recently, \cite{Lakshmana2016} proposed a conservative precoder design  with the objective of maximizing the weighted sum-rate of UEs for arbitrary UE-centric clustering method with incomplete  CSIs, where the long term channel statistic was incorporated into the optimization. Finally, \cite{Liu2016tvt} designed a clustering scheme maximizing the average net throughput of the dense C-RAN by taking the training overhead into account. The scheme is based on a hybrid CoMP transmission mode and operates under a long time duration that may be performed at the medium access control (MAC) layer  since only large-scale CSIs are required. However, both the beam directions and power allocations were not optimized in \cite{Liu2016tvt}. None of the papers \cite{Shi2014ICC,Kim2014,Lakshmana2016,Liu2016tvt} considered  the fronthaul capacity constraints  and were mainly focused on sum-rate maximization problems without incorporating QoS requirements.

The aim of this paper is to provide a complete framework to jointly tackle the above-mentioned  challenges together. Specifically, we investigate the joint optimization of  RRH selection, RRH-UE associations and transmit beamforming to minimize the NPC for downlink multi-channel C-RAN with incomplete CSI, subject to fronthaul link capacity constraints, all UEs' rate requirements and per-RRH power constraints. The NPC is modeled as the sum of the RRH power consumption and the fronthaul link power consumption. The low-power sleep mode is considered in the RRH power consumption model, and the fronthaul link power consumption is modeled as a linear function of fronthaul traffic. To reduce the computational complexity, each UE is restricted to be served by its nearby RRHs since only nearby RRHs contribute significantly to the UE's signals. Moreover, to reduce the channel measurement overhead, we introduce the subset of RRHs that each UE should estimate the CSIs to, while the large-scale fading (such as path-loss and shadowing) is assumed to be known for the other unavailable CSI. In general, the candidate set of RRHs for serving UEs and the CSI estimation set of RRHs for each UE are determined based on UEs' locations that may be the task of the upper-layer, which is beyond the scope of this paper. The NPC minimization problem is an NP-hard mixed-integer non-linear programming (MINLP) problem due to the indicator functions introduced in both objective function and fronthaul capacity constraints, whose optimal solution is intractable. In addition, due to the sum rate constraints and incomplete CSI, the QoS constraints are non-convex and difficult to handle. Furthermore, due to the conflicting constraints, the NPC minimization problem may be infeasible and the initialization solution should be carefully selected. As a result, the contributions of this paper can be summarized as follows:
\begin{enumerate}
  \item Due to the incomplete CSI, it is intractable to derive the exact closed-form expression of the data rate for each UE, and thus stringent QoS requirements for each UE are difficult to be guaranteed. To alleviate this difficulty, we conservatively replace the data rate of each UE with its lower-bound expression derived by using the Jensen's inequality.
  \item  To resolve the feasibility issue, we provide a low-complexity UE selection algorithm based on bisection search method to maximize the number of admitted UEs that can achieve their QoS targets, and its complexity  only  increases logarithmically with the number of UEs. Simulation results show that this algorithm can achieve marginal performance loss with respect to (w.r.t.) that obtained by the exhaustive UE search algorithm with an exponential computational complexity over the number of admitted UEs.
  \item Given the feasible set of UEs from the UE selection algorithm, we provide a low-complexity single-layer iterative algorithm (i.e., Algorithm 1) to solve the NPC minimization problem. Specifically, the non-smooth indicator function is approximated as a non-convex function and  the successive convex approximation (SCA) technique \cite{beck2010sequential} is adopted to approximate the non-convex function as a series of convex functions. To deal with the non-convex QoS constraints, we translate the technique in \cite{Qingjiang2011} that aimed at rate maximization problem to the NPC minimization problem with rate expressions in the constraints and incomplete CSI. The convergence of the iterative algorithm is strictly proved.
  \item In each iteration of Algorithm 1, there is a subproblem that the beam-vectors should be optimized. We derive the structure of the optimal beam-vectors by employing the Lagrange dual decomposition method. Then, each beam-vector can be obtained in parallel for each sub-channel (SC), which facilitates  the application of the cloud computing technique in BBU pool.
\end{enumerate}

This paper is organized as follows. Section \ref{system} presents the system model, and Section \ref{probleforana} formulates the UE selection problem and  NPC minimization problem along with the complexity analysis. The single-layer iterative algorithm to solve the NPC algorithm is given in Section \ref{NPCalg} when the UEs are selected to be admitted. Then, in Section \ref{userselectionalg}, the low-complexity UE selection algorithm is provided. Simulation results are presented in Section \ref{userselectionalg} to evaluate the performance of the proposed algorithms. Finally, conclusions are drawn in Section \ref{conclu}.



Notations: For a set ${\cal A}$, $\left| {\cal A} \right|$ denotes the cardinality of ${\cal A}$, while for a complex number $x$, $\left| x \right|$ denotes   the magnitude of $x$. $\bf{1}$ denotes a vector with all elements equal to ones. `s.t.' is short for `subject to'. ${{\mathbb{E}}_{\{ x\} }}\{ y\} $ means the expectation of $y$ over $x$. The complex Gaussian distribution is denoted as ${\cal C}{\cal N}(\cdot, \cdot)$. We use ${\mathbb{ C}}$ to represent the complex set. The lower-case bold letters denote  vectors and upper-case bold letters denote matrices. ${\rm{blkdiag}}(\cdot)$ denotes the block diagonalization operation.

\section{System Model}\label{system}

\subsection{System model}

\begin{figure}
\centering
\includegraphics[width=4.5in]{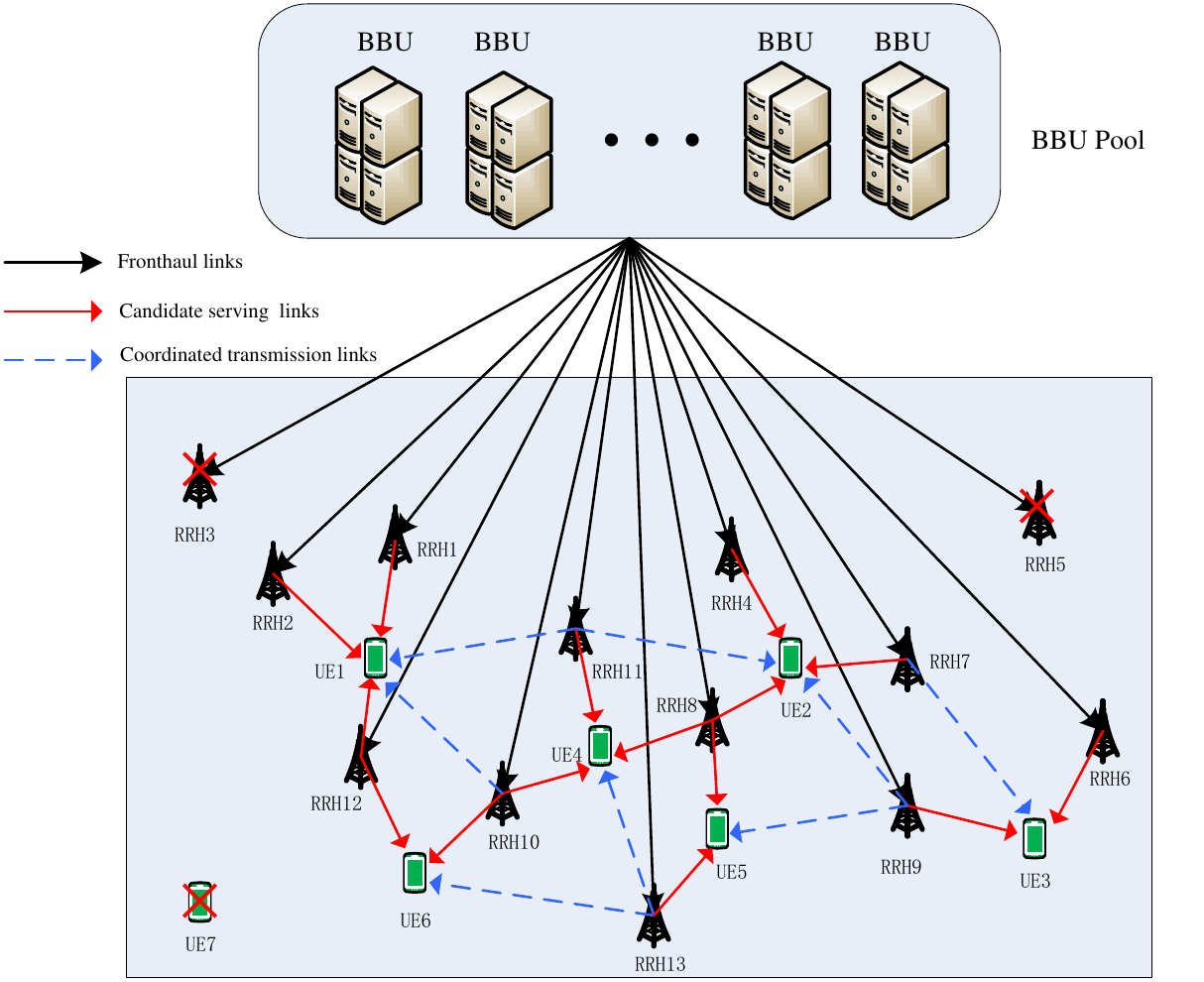}
\caption{Illustration of a C-RAN with thirteen RRHs and seven UEs. The RRHs are connected to a BBU pool through wireless fronthaul links. In this scenario,
UE 7 is not selected for serving and the candidate sets of RRHs for the selected UEs are given by  ${{\cal I}_1} = \{1, 2,12\} $, ${{\cal I}_2} = \{4,7,8\} $, ${{\cal I}_3} = \{6,9\} $, ${{\cal I}_4} = \{8,10,11\} $, ${{\cal I}_5} = \{8, 13\} $ and ${{\cal I}_6} = \{10, 12\} $, respectively. The sets of UEs that are potentially served by the RRHs are given by ${{ {\cal U}}_1}=\{1\}$, ${{ {\cal U}}_2}=\{1\}$, ${{ {\cal U}}_3}=\{\emptyset\}$, ${{ {\cal U}}_4}=\{2\}$, ${{ {\cal U}}_5}=\{\emptyset\}$, ${{ {\cal U}}_6}=\{3\}$, ${{ {\cal U}}_7}=\{2\}$, ${{ {\cal U}}_8}=\{2,4,5\}$, ${{ {\cal U}}_9}=\{3\}$, ${{ {\cal U}}_{10}}=\{4,6\}$, ${{ {\cal U}}_{11}}=\{4\}$, ${{ {\cal U}}_{12}}=\{1,6\}$ and ${{ {\cal U}}_{13}}=\{5\}$, respectively. The sets of RRHs for coordinating the interference for the selected UEs are given by ${\cal C}{_1}=\{10,11\}$, ${\cal C}{_2}=\{9,11\}$, ${\cal C}{_3}=\{7\}$, ${\cal C}{_4}=\{13\}$, ${\cal C}{_5}=\{9\}$ and ${\cal C}{_6}=\{13\}$, respectively. The sets of coordinated UEs by the RRHs are given by  ${{ {\cal T}}_7}=\{3\}$,  ${{ {\cal T}}_9}=\{2,5\}$, ${{ {\cal T}}_{10}}=\{1\}$, ${{ {\cal T}}_{11}}=\{1,2\}$  and ${{ {\cal T}}_{13}}=\{4,6\}$, respectively. }
\label{fig1}
\end{figure}

Consider a downlink ultra-dense C-RAN, as shown in Fig.~\ref{fig1}, consisting of $I$ RRHs and $K$ UEs\footnote{ Due to the simple functionalities of RRHs, the RRHs can be densely deployed with low hardware cost, wherein the number of RRHs may even be larger than that of UEs. Hence, the average distance between RRHs and UEs can be significantly reduced. As a result, the transmission power of the RRHs can also be reduced due to the decreased path loss. }, where each RRH is equipped with $M$ transmit antennas, and each UE has a single antenna. Denote the set of RRHs and UEs as ${\cal I} = \left\{ {1, \cdots ,I} \right\}$ and $\bar {\cal U} = \left\{ {1, \cdots ,K} \right\}$, respectively. Each RRH is connected to the BBU pool through wireless (e.g. mmWave communication)  fronthaul links. The  fronthaul links are represented by dark solid arrows in Fig.~\ref{fig1}. The BBU pool is assumed to have all UEs' data and distributes each UE's data to a carefully selected set of RRHs through the  fronthaul links. It is assumed that all the RRHs send their received data using the Orthogonal Frequency Division Multiple Access (OFDMA) technique and then cooperatively transmit to the UEs.

Denote ${ {\cal U}} \subseteq \bar {\cal U}$ as the subset of UEs that are admitted in the C-RAN. To reduce the computational complexity of the large network, it is assumed that each UE $k\in {\cal U} $ can only be served by its nearby RRHs since only nearby RRHs contribute significantly to the UE's signal quality due to the severe path loss. Denote ${{\cal I}_k} \subseteq {\cal I}$ and ${{ {\cal U}}_i}\subseteq { {\cal U}} $ as the candidate set of RRHs that potentially serve UE $k$ and the set of UEs that can be potentially served by RRH $i$, respectively. The transmission links from the RRHs in ${{\cal I}_k}$ to UE $k$ are called the candidate serving links, which are represented in red solid arrows in Fig.~\ref{fig1}. In this paper, it is assumed that ${{\cal I}_k} $ and ${{ {\cal U}}_i}$ are predetermined by some well-known user-centric cluster methods \cite{Penggenwcl2014,Papadogiannis,Liu2016tvt} determined by the MAC layer\footnote{In general, the cluster method is mainly determined based on the large-scale CSI, which is usually performed in the upper layer such as MAC layer. In some hot spots such as stadia and shopping malls, the users move slowly. Hence, the cluster can be kept fixed for a long time compared with the instantaneous CSI. This paper only focuses on the beam-vectors at the physical layer, and how to design the optimal cluster method is beyond the scope of this paper.}. Please refer to \cite{Bassoy2017} for a survey on user-centric cluster methods. Note that since no restrictions are placed on ${{\cal I}_k} $, they can overlap with each other, i.e., there may exist two different UEs $k$ and $k'$ that ${{\cal I}_k}\cap {{\cal I}_{k'}}\ne \emptyset$, for  $\forall k,k'\in { {\cal U}}$. Moreover, the other-cluster interference due to overlapping coverage can be effectively handled under this user-centric cluster method. For example, UE 4 and UE 5 have one common serving RRH 8. Hence, RRH 8 will transmit useful signals to both UE 4 and UE 5, rather than only interference signals. In addition, the BBU pool has the CSI knowledge from RRH 3 to UE 4. Thus, the interference from RRH 3 to UE 4 will be carefully controlled when RRH 3 is serving UE 5. In contrast to the non-cooperative optimization where each cluster selfishly optimizes its own performance without considering its impact on the other clusters, in dense C-RAN all the signal processing operation is performed at the BBU pool, where the interference among different clusters can be centrally mitigated by resorting to the powerful cloud computing tool.

Denote the set of available sub-channels (SCs) as ${\cal N}= \{ 1,2, \cdots ,N\} $, where $N$ is the total number of SCs. To maximize the spectral efficiency, it is assumed that universal frequency reuse is adopted and the multiuser interference can be efficiently handled by the beamforming technique. Denoting ${\bf{w}}_{i,k}^{(n)}\in {{\mathbb{ C}}^{M \times 1}}$ as the beam-vector at RRH $i$ for UE $k$ on SC $n$, the transmitted signal of RRH $i$ on SC $n$ is
\begin{equation}\label{trnasmitsignal}
  {\bf{x}}_i^{(n)} = \sum\nolimits_{k \in {{\cal U}_i}} {{\bf{w}}_{i,k}^{(n)}s_{k}^{(n)}},
\end{equation}
where ${s_k^{(n)}}$ is the data symbol for UE $k$ on SC $n$. Without  loss of generality,  it is assumed that  ${{\mathbb{E}}}\{ |s_k^{(n)}{|^2}\}  = 1$ and ${{\mathbb{E}}}\{ s_{{k_1}}^{({n_1})}s_{{k_2}}^{({n_2})}\}  = 0$ for $({n_1},{k_1}) \ne ({n_2},{k_2}),\forall {n_1},{n_2} \in {\cal N},\forall {k_1},{k_2} \in {\cal U}$. The baseband received signal  at UE $k$ on SC $n$ is given by
\begin{equation}\label{receivedsignal}
  y_k^{(n)} = \underbrace {\sum\nolimits_{i \in {{\cal I}_k}} {{\bf{h}}_{i,k}^{(n)}{\bf{w}}_{i,k}^{(n)}s_k^{(n)}} }_{{\rm{desired\  signal}}} + \underbrace {\sum\nolimits_{l \ne k,l \in {\cal U}} {\sum\nolimits_{i \in {{\cal I}_l}} {{\bf{h}}_{i,k}^{(n)}{\bf{w}}_{i,l}^{(n)}s_l^{(n)}} } }_{ {\rm{interference}}} + z_k^{(n)},
\end{equation}
where ${\bf{h}}_{i,k}^{(n)} \in {{\mathbb{ C}}^{1 \times M}}$ is the channel vector from RRH $i$ to UE $k$ on SC $n$, and $z_k^{(n)}$ is the additive complex white Gaussian noise following the distribution of ${\cal C}{\cal N}(0,\sigma _k^2)$. The channel vector ${\bf{h}}_{i,k}^{(n)}$ can be written as ${\bf{h}}_{i,k}^{(n)} = \alpha _{i,k}^{(n)}{\bf{\tilde h}}_{i,k}^{(n)}$, where $\alpha _{i,k}^{(n)}$ denotes the large-scale channel gain that includes the path loss and shadowing, and ${\bf{\tilde h}}_{i,k}^{(n)}$ denotes the small-scale fading vector, where all elements are dependent of each other and  each one has zero mean and unit variance.

For the sake of reduced complexity of decoding at the receivers, we do not consider the joint decoding of the interfering signals and the multiuser interference is simply regarded as noise at the receivers. In addition, coherent joint transmission\footnote{This assumption is valid for dense C-RAN. The reason is that dense C-RAN  is usually deployed in hot spots with smaller coverage area compared with that of the conventional C-RAN that covers multiple macrocells \cite{mobile2011c}. Hence, different transmission delays due to different transmission distances between RRHs and BBU pool can be ignored. Then, both the synchronization and coherent joint transmission are possible.} is assumed as in most of existing papers \cite{Yuanming2014,Shixin2015,dai2016energy,PAN17,Jian2013,Binbin2014,dwkNg2015,Liu2016,Ha2014,Abdelnasser2016}. Then, the  SINR at UE $k$ on SC $n$ can be obtained from (\ref{receivedsignal}) as
\begin{equation}\label{sinr}
 \gamma _k^{(n)}({\bf{w}}) = \frac{{{{\left| {\sum\nolimits_{i \in {{\cal I}_k}} {{\bf{h}}_{i,k}^{(n)}{\bf{w}}_{i,k}^{(n)}} } \right|}^2}}}{{\sum\nolimits_{l \ne k,l \in {\cal U}} {{{\left| {\sum\nolimits_{i \in {{\cal I}_l}} {{\bf{h}}_{i,k}^{(n)}{\bf{w}}_{i,l}^{(n)}} } \right|}^2}}  + \sigma _k^2}}.
\end{equation}
where ${\bf{w}}$ denotes the collection of all beam-vectors.

As seen in (\ref{sinr}), to design the beam-vectors for all UEs, the overall CSI of all UEs is required. However, it is a formidable task to obtain all CSI for the dense C-RAN due to the limited training resources. To handle this difficulty, we introduce the set ${\widetilde {\cal I}_k}  \supseteq  {{\cal I}_k}$ for each UE $k$ that is defined as the set of RRHs that UE $k$ needs to measure CSI from. Also, we define ${\widetilde {\cal U}_i}  \supseteq  {{\cal U}_i}$ for each RRH $i$ as the set of UEs that each RRH $i$ knows the CSI to. In general, ${\widetilde {\cal I}_k}$ are the set of UE $k$'s nearby RRHs and ${\widetilde {\cal U}_i}$ are the set of RRH $i$'s nearby UEs.  Note that at least the CSI from all RRHs in ${{\cal I}_k}$  is required for cooperative transmission design. The other CSI from RRHs in ${\cal C}{_k} = {\widetilde {\cal I}_k}\backslash {{\cal I}_k}$ to UE $k$ is used to coordinate the interference, and the links from RRHs in ${\cal C}{_k}$ are called coordinated interference links, which are shown by blue dashed arrows in Fig.~\ref{fig1}. Also, the UEs in ${\cal T}_i={\widetilde {\cal U}_i}  \backslash  {{\cal U}_i}$ are called RRH $i$'s coordinated UEs. For the CSI from RRHs in ${\cal I}\backslash {\widetilde {\cal I}_k}$ to UE $k$, it is assumed that the BBU pool only knows the large scale gains $\{\alpha _{i,k}^{(n)}, \forall i\in {\cal I}\backslash {\widetilde {\cal I}_k},k\in {\cal U},n \in \cal N\}$. This is possible because the large scale gains change much more slowly than the small-scale fading.

Since the CSI in ${\cal I}\backslash {\widetilde {\cal I}_k}$ is unknown, we consider the following data rate for UE $k$ on SC $n$ (bit/s/Hz) \cite{cover2012elements}
\begin{equation}\label{averagerate}
  {\bar r}_k^{(n)}({\bf{w}}) = {{\mathbb{E}}_{\left\{ {{\bf{h}}_{i,k}^{(n)},i \in {\cal I}\backslash {{\widetilde {\cal I}}_k}} \right\}}}\left\{ {{{\log }_2}(1 + \gamma _k^{(n)}({\bf{w}}))} \right\}.
\end{equation}
where the expectation operator is performed over the fast fading of the unknown CSI in ${\cal I}\backslash {\widetilde {\cal I}_k}$.
Each UE $k$'s total data rate should be larger than the minimum rate requirement ${R_{k,\min }}$:
\begin{equation}\label{ratecons}
  {\rm{C1:}}\  {\bar r}_{k,\rm{tot}}({\bf{w}})=\sum\nolimits_{n \in {\cal N}} {{\bar r}_k^{(n)}({\bf{w}})}  \ge {R_{k,\min }},\forall k \in {\cal U}.
\end{equation}

In each fronthaul link, the maximum capacity that can be supported is limited. Hence, the following fronthaul capacity constraint follows:
\begin{equation}\label{fronthaulca}
  {\rm{C2}:}\ \sum\nolimits_{k \in {\cal U}_i} {\varepsilon \left( P_{i,k}^{{\rm{tr}}}({\bf{w}}) \right){{\bar r}_{k,{\rm{tot}}}({\bf{w}})}}  \le {C_{i,\max }},\forall i \in {\cal I},
\end{equation}
where  $\varepsilon \left( \cdot \right)$ is an indicator function,  defined as
\begin{equation}\label{indifunc}
  \varepsilon \left( x \right) = \left\{ {\begin{array}{*{20}{l}}
{1,\;{\rm{if}}\;x \ne 0,}\\
{0,\;{\rm{otherwise}},}
\end{array}} \right.
\end{equation}
$P_{i,k}^{{\rm{tr}}}({\bf{w}}) = \sum\nolimits_{n \in {\cal N}} {{{\left\| {{\bf{w}}_{i,k}^{(n)}} \right\|}^2}} $ denotes the total transmission power from RRH $i$ to UE $k$, $C_{i,\max }$ is the maximum capacity that can be supported by the $i$th fronthaul link.

\subsection{Network power consumption model}
In this subsection,  a practical NPC model is provided that consists of two parts: power consumption at the RRHs and power consumption on the fronthaul links.

As in \cite{Auer2011}, the power consumption of RRH $i$ can be modeled as a piecewise linear function of the transmit power at RRH $i$:
\begin{equation}\label{rrhpower}
  P_i^{{\rm{rrh}}}({\bf{w}})= \left\{ \begin{array}{l}
{{{\eta _i}}}P_i^{{\rm{tr}}}({\bf{w}})+ P_i^{{\rm{active}}},\ {\rm{if  }}\ P_i^{{\rm{tr}}}({\bf{w}}) > 0\\
P_i^{{\rm{sleep}}},\qquad \qquad\quad\  \   {\rm{if}}\ P_i^{{\rm{tr}}}({\bf{w}}) = 0
\end{array} \right.
\end{equation}
where $\eta _i>1$ is the constant accounting for the  efficiency of the power amplifier of RRH $i$, $P_i^{{\rm{tr}}}({\bf{w}})$ is the total transmit power at RRH $i$ that should be no larger than  ${P_{i,\max }}$, i.e.,
\begin{equation}\label{powercons}
 {\rm{C3}}:\ P_i^{{\rm{tr}}}({\bf{w}}){\rm{ = }}\sum\nolimits_{k \in {{\cal U}_i}} {P_{i,k}^{{\rm{tr}}}({\bf{w}})}   \le {P_{i,\max }},i \in {\cal I},
\end{equation}
$P_i^{{\rm{active}}}$ and $P_i^{{\rm{sleep}}}$ represent the circuit power consumption when RRH $i$ is in active mode and sleep mode, respectively. In general, $P_i^{{\rm{active}}}$ is much larger than $P_i^{{\rm{sleep}}}$, which motivates us strategically to switch off the RRHs to save power in case of very low traffic.

Fronthaul power consumption model is critical for the optimization of NPC. In \cite{Yuanming2014} and \cite{Shixin2015}, the fronthaul power consumption was simply modeled as a step function, with a larger constant value for active mode and smaller one for sleep mode. In \cite{Yong2013}, the fronthaul power consumption is modeled to be proportional to the number of UEs that each one supports. However, these papers did not take into account the effect of data rate transmitting on each fronthaul link. Intuitively, to support high fronthaul transmit data rate, more power should be consumed on the fronthaul links. Compared with \cite{Yuanming2014,Shixin2015,Yong2013}, we go one step further by modeling the power consumption of each fronthaul link to be proportional to the total fronthaul transmit data rate as in \cite{Fehske2010}:
\begin{equation}\label{frpower}
P_i^{{\rm{fr}}}({\bf{w}}) = {\rho _i} \sum\nolimits_{k \in {\cal U}_i} {\varepsilon \left( {P_{i,k}^{{\rm{tr}}}({\bf{w}})} \right){{\bar r}_{k,{\rm{tot}}}({\bf{w}})}},
\end{equation}
where ${\rho _i}$ is a constant scaling factor \footnote{In general, this scaling factor may not be a constant, rather depend on the total transmit data rate on the fronthaul link. However, how to accurately model this relationship is still under investigation. To the best of our knowledge, only \cite{Fehske2010} provided the detailed study of this model that has been adopted by the existing work such as \cite{dai2016energy}.}.

Based on the above analysis and with some simple manipulations, the NPC is modeled as
\begin{eqnarray}
{P_{{\rm{NPC}}}}({\bf{w}})\!\!\! &=&\!\!\! \sum\limits_{i \in {{\cal I}}} {\left\{ {P_i^{{\rm{rrh}}}({\bf{w}}) + P_i^{{\rm{fr}}}}({\bf{w}}) \right\}}   \nonumber\\
 \!\!\!&=& \!\!\!\sum\limits_{i \in {{\cal I}}} {\left\{ {{{{{\eta _i}}}{P_i^{{\rm{tr}}}({\bf{w}})}} \!+\! \varepsilon \left(P_i^{{\rm{tr}}}({\bf{w}}) \right)P_i^{\rm{c}}\!\! + {\rho _i} \sum\limits_{k \in {\cal U}_i} {\varepsilon \left( {P_{i,k}^{{\rm{tr}}}({\bf{w}})} \right){{\bar r}_{k,{\rm{tot}}}({\bf{w}})}}} \right\}}  \!+\! \sum\limits_{i \in {{\cal I}}} {P_i^{{\rm{sleep}}}} ,\label{totalpow}
\end{eqnarray}
where ${P_i^{{\rm{tr}}}}({\bf{w}})$ is given in (\ref{powercons}), $P_i^{\rm{c}} = P_i^{{\rm{active}}} - P_i^{{\rm{sleep}}}, \forall i\in {\cal I}$.

\section{Problem Formulation and Analysis}\label{probleforana}

Based on the above system model, we formulate the user selection problem and the NPC minimization problem in a two-stage form. Then, we provide the complexity analysis for the formulated problems.

\subsection{Problem Formulation}\label{djfore}

Due to the limited fronthaul capacity constraints C2 in (\ref{fronthaulca}) and the power constraints  C3 in (\ref{powercons}), the system may not be able to support all UEs with their rate requirements of C1 in (\ref{ratecons}). Hence, some UEs may be dropped or rescheduled in other orthogonal time slots to make the optimization problem feasible. As a result, we may consider a two-stage optimization problem. In the first stage, one should find the largest subsets of UEs that can be supported by the system\footnote{Dense C-RAN is usually deployed in hot spots (such as shopping mall, stadia, et al.) where the number of UEs is huge, and the amount of available communication resource is limited. Hence,  maximizing the number of admitted users for each time slot should be placed in higher priority. In some other scenarios, where there are abundant wireless resource, maximizing the number of admitted UEs in each time slot may not be a good option in reducing NPC, and dynamically scheduling the UE in different time slots may further reduce NPC, which will be left for future work. }, while in the second stage, one should optimize the corresponding beam-vectors to minimize $P_{{\rm{NPC}}}$ with the selected subset of UEs obtained from the first stage.

As a result, the optimization problem at the first stage is formulated as
\begin{equation}\label{staone}
\begin{array}{l}
 {\cal P}_1:\ \mathop {\max }\limits_{{\bf{w}}, {\cal U} \subseteq {\overline {\cal U}} } \quad \left| \cal U \right|\\
\qquad\ {\rm{s}}.{\rm{t}}.\qquad {\kern 1pt} {\rm{C1}},{\rm{C2}},{\rm{C3}}.
\end{array}
\end{equation}
Denote ${{\cal U}^\star}$ as the solution from Stage I and the corresponding ${{ {\cal U}}_i}$ becomes ${{ {\cal U}}_i^\star}$. Then, the optimization problem at the second stage is formulated as
\begin{subequations}\label{mainpro}
\begin{align}
{\cal P}_2:\ \mathop {\min }\limits_{{\bf{w}}} \quad
& \sum\nolimits_{i \in {{\cal I}}} {\left\{ {{{{{\eta _i}}}{P_i^{{\rm{tr}}}({\bf{w}})}} \!+\! \varepsilon \left(P_i^{{\rm{tr}}}({\bf{w}}) \right)P_i^{\rm{c}}\!\! + {\rho _i} \sum\nolimits_{k \in {\cal U}_i^\star} {\varepsilon \left( {P_{i,k}^{{\rm{tr}}}({\bf{w}})} \right){{\bar r}_{k,{\rm{tot}}}({\bf{w}})}}} \right\}}  \label{objfunct}
\\
\qquad\ \textrm{s.t.}\qquad\!\!\!\!
&{\rm{C1}},{\rm{C2}},{\rm{C3}}
\end{align}
\end{subequations}
In the constraints C1, C2, and C3, ${{\cal U}}$ and ${{ {\cal U}}_i}$ are replaced by ${{\cal U}^\star}$ and ${{ {\cal U}}_i^\star}$, respectively. Note that the constant term $\sum\nolimits_{i \in {{\cal I}}} {P_i^{{\rm{sleep}}}} $ in (\ref{totalpow}) has been omitted in the objective function (\ref{objfunct}).

We emphasize that the aim of Stage I is to find the maximum number of admitted UEs with feasible beam-vectors. These obtained beam-vectors are not guaranteed to be optimal in terms of NPC. Hence, we need to perform Stage II to  optimize the beam-vectors to reduce the NPC. The beam-vectors obtained from Stage I will be a feasible initial input that is required by the algorithm developed in Stage II.

The incomplete CSI at the BBU pool makes the design of beam-vectors very difficult to solve and the expression for the  data rate is difficult to derive. In the following, we consider its lower-bound and replace the data rate with its lower-bound, which makes the optimization problem more tractable.

We first simplify the SINR expression in (\ref{sinr}). The beam-vectors for each UE on each SC $n$ are merged into a single large-dimension vector $ { {{\bf{\bar w}}}}_k^{(n)} = {[ {{\bf{w}}_{i,k}^{(n){\rm{H}}},\forall i \in {{\cal I}_k}} ]^{\rm{H}}} \in {{\mathbb{C}}^{\left| {{{\cal I}_k}} \right|M \times 1}}, \forall n\in \cal N$. Then, we define a set of new channel vectors ${\bf{\bar h}}_{l,k}^{(n)} = [ {{\bf{h}}_{i,k}^{(n)},\forall i \in {{\cal I}_l}} ] \in {{\mathbb C}^{1 \times \left| {{{\cal I}_l}} \right|M}}$, representing the aggregated CSI from the RRHs in ${\cal I}_l$ to UE $k$ on SC $n$. The SINR expression in (\ref{sinr}) can be rewritten as
\begin{equation}\label{sinrre}
  \gamma _k^{(n)}{({\bf{w}})} = \frac{{{{\left| {{\bf{\bar h}}_{k,k}^{(n)}{\bf{\bar w}}_k^{(n)}} \right|}^2}}}{{\sum\nolimits_{l \ne k,l \in {\cal U}} {{{\left| {{\bf{\bar h}}_{l,k}^{(n)}{\bf{\bar w}}_l^{(n)}} \right|}^2}}  + \sigma _k^2}}.
\end{equation}
Note that  ${{\bf{\bar h}}_{k,k}^{(n)}}$ is perfectly known in the BBU pool according to the previous assumption, and only the denominator in (\ref{sinrre}) contains the uncertain terms. However, it is difficult to obtain the accurate rate expression. To deal with this challenge, we consider its lower-bound with more tractable form. Specifically, since ${\log _2}\left( {1 + {a \mathord{\left/
 {\vphantom {1 x}} \right.
 \kern-\nulldelimiterspace} x}} \right)$ is a convex function for any positive $a$, by using Jensen's inequality \cite{boyd2004convex}, the lower bound of the data rate in (\ref{averagerate}) can be derived as
 \begin{eqnarray}
 &&\bar r_k^{(n)}{({\bf{w}})}\\
  &\ge& {\log _2}\left( {1 + \frac{{{{\left| {{\bf{\bar h}}_{k,k}^{(n)}{\bf{\bar w}}_k^{(n)}} \right|}^2}}}{{{{\mathbb{E}}_{\left\{ {{\bf{h}}_{i,k}^{(n)},i \in {\cal I}\backslash {{\tilde {\cal I}}_k}} \right\}}}\left\{ {\sum\nolimits_{l \ne k,l \in {\cal U}} {{{\left| {{\bf{\bar h}}_{l,k}^{(n)}{\bf{\bar w}}_l^{(n)}} \right|}^2}} } \right\} + \sigma _k^2}}} \right)\label{firstineq}\\
 &=& {\log _2}\left( {1 + \frac{{{{\left| {{\bf{\bar h}}_{k,k}^{(n)}{\bf{\bar w}}_k^{(n)}} \right|}^2}}}{{\sum\nolimits_{l \ne k,l \in {\cal U}} {{\bf{\bar w}}_l^{(n){\rm{H}}}{\bf{A}}_{l,k}^{(n)}{\bf{\bar w}}_l^{(n)}}  + \sigma _k^2}}} \right)\label{secondg}\\
 &\buildrel \Delta \over =&  \tilde r_k^{(n)}{({\bf{w}})}\label{thirdp}
 \end{eqnarray}
where ${\bf{A}}_{l,k}^{(n)} = {{\mathbb{E}}_{\left\{ {{\bf{h}}_{i,k}^{(n)},i \in {\cal I}\backslash {{\tilde {\cal I}}_k}} \right\}}}\left\{ {{\bf{\bar h}}{{_{l,k}^{(n)}}^{\rm{H}}}{\bf{\bar h}}_{l,k}^{(n)}} \right\}\in {\mathbb{C}}^{M{|{\cal I}_l|} \times M{|{\cal I}_l|}}$. To obtain the closed-form expression of ${\bf{A}}_{l,k}^{(n)} $, we define the indices of ${{{\cal I}_l}}$ as ${{\cal I}_l} = \{ s_1^l, \cdots ,s_{|{{\cal I}_l}|}^l\} $. Then, we have
\begin{equation}\label{akk}
{\bf{A}}_{l,k}^{(n)} = \left[ {\begin{array}{*{20}{c}}
{{{\left( {{\bf{A}}_{l,k}^{(n)}} \right)}_{1,1}}}& \cdots &{{{\left( {{\bf{A}}_{l,k}^{(n)}} \right)}_{1,|{{\cal I}_l}|}}}\\
 \vdots & \ddots & \vdots \\
{{{\left( {{\bf{A}}_{l,k}^{(n)}} \right)}_{|{{\cal I}_l}|,1}}}& \cdots &{{{\left( {{\bf{A}}_{l,k}^{(n)}} \right)}_{|{{\cal I}_l}|,|{{\cal I}_l}|}}}
\end{array}} \right],l\neq k
\end{equation}
where ${\left( {{\bf{A}}_{l,k}^{(n)}} \right)_{i,j}} \in {{\mathbb{C}}^{M \times M}},i,j \in 1, \cdots ,|{{\cal I}_l}|$  is the block matrix of ${\bf{A}}_{l,k}^{(n)}$ at the $i$th row and $j$th column,   given by
\begin{equation}\label{ak}
 {\left( {{\bf{A}}_{l,k}^{(n)}} \right)_{i,j}} = \left\{ \begin{array}{l}
{\bf{h}}_{s_i^l,k}^{(n){\rm{H}}}{\bf{h}}_{s_j^l,k}^{(n)},\ \ \ \ \ {\rm{ if }}\ s_i^l,s_j^l \in {{\tilde {\cal I}}_k},\\
\left|\alpha _{s_i^l,k}^{(n)}\right|^2{{\bf{I}}_{M \times M}},\ {\rm{if}}\ s_i^l,s_j^l \notin {{\tilde {\cal I}}_k},{\rm{and }}\ i = j,\\
{{\bf{0}}_{M \times M,}}\qquad\ \  \ \ {\rm{otherwise.}}
\end{array} \right.
\end{equation}
It can be easily verified that ${\bf{A}}_{l,k}^{(n)}$ is a positive definite matrix. Note that the derivations of matrix  ${\bf{A}}_{l,k}^{(n)}$ place no restrictions on the channel distributions and only large-scale channel gains are required. Hence, the following developed algorithms are applicable for any channel distributions, such as Rayleigh fading, Ricean channels, Nakagami-$m$ fading channels, et al.

\begin{figure}
\begin{minipage}[t]{0.485\linewidth}
\centering
\includegraphics[width=2.8in]{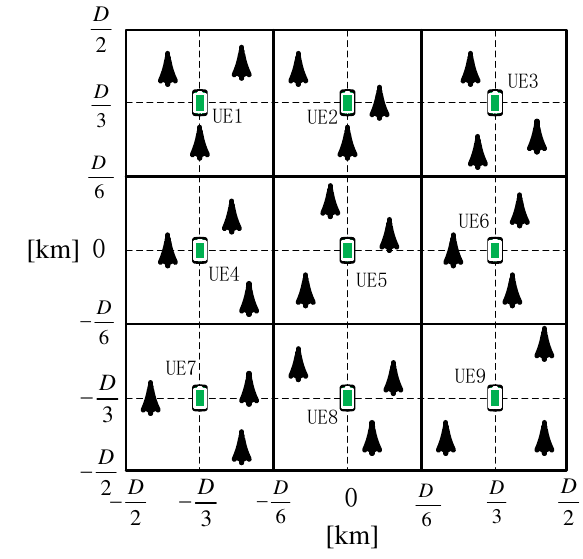}
\caption{Non-overlapped cluster topology.}
\label{simulsetup}
\end{minipage}%
\hfill
\begin{minipage}[t]{0.485\linewidth}
\centering
\includegraphics[width=3.1in]{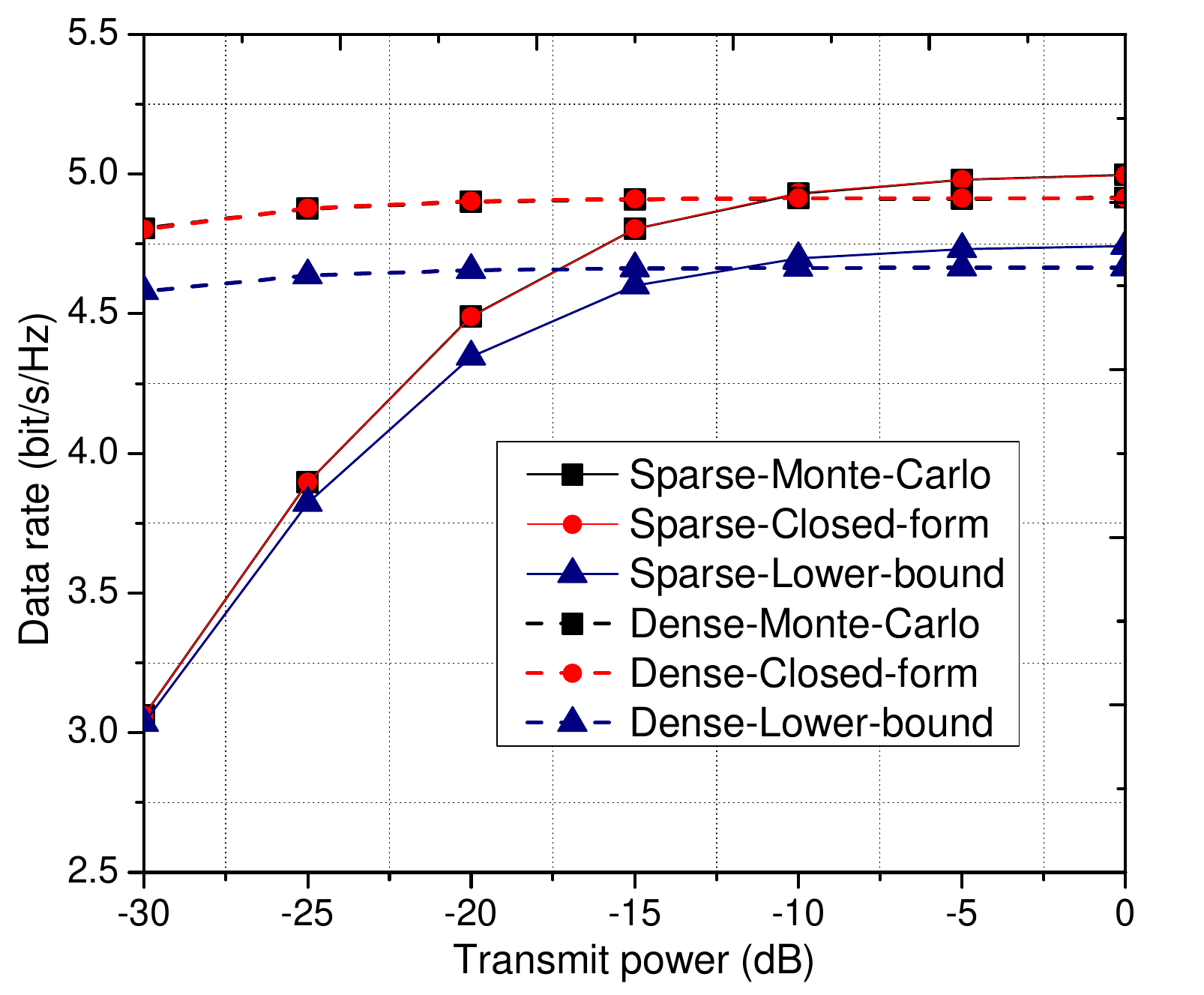}
\caption{Data rate versus the transmit power for the special case.}
\label{tightnesspic}
\end{minipage}
\end{figure}

We now start to check the tightness of this rate lower-bound. It is difficult to derive the accurate data rate expression for general case. Instead, in Appendix \ref{specialcase}, we derive the accurate closed-form expression of data rate for one special case under three assumptions: 1) The RRH serving cluster is the same as the CSI cluster for each UE: ${\cal I}_k={\widetilde {\cal I}_k}$; 2) The RRH serving cluster for each UE is non-overlapped with each other: ${{\cal I}_k}\cap {{\cal I}_{k'}}= \emptyset, \forall k,k'\in \cal U$; 3) The small-scale fading vector ${\bf{\tilde h}}_{i,k}$ follows the distribution of ${\cal C}{\cal N}({\bf{0}},{\bf{I}})$ for $\forall i,k$. We consider one non-overlapped C-RAN scenario deployed within a square area of coordinates $[-D/2, D/2]\times [-D/2, D/2]$ km as shown in Fig.~\ref{simulsetup}. This network area is divided into nine $D/3\  {\rm{km}} \times D/3\ {\rm{km}}$ squares. In each square, one UE is located at the center point and three RRHs are randomly generated in this square to exclusively serve this UE. For simplicity, only one SC is considered. The other simulation parameters are the same as in the simulation Section. It is assumed that each RRH transmits at their maximum power and the beam direction is chosen to be channel direction. The values of $D=3$ and $D=1$ are tested, which correspond to sparse and dense scenarios, respectively. Only UE 5 is considered.  Fig.~\ref{tightnesspic} plots three kinds of curves for comparison: one is the lower bound of data rate derived in (\ref{thirdp}), one is the accurate closed-form data rate expression derived in (\ref{finalexpre}) in Appendix \ref{specialcase}, and the last one is the Monte-Carlo simulations. It is seen from Fig.~\ref{tightnesspic} that the curve of the closed-form expression coincides with that of the Monte-Carlo simulations, which verifies the correctness of the derivations.
Furthermore, for the sparse scenarios, when the transmit power is low, $P_{\rm{max}}<-20 {\rm{dB}}$, the lower-bound is quite tight. With the increase of transmit power, the gap increases and becomes a constant in the high transmit power regime. Note that only roughly 3\% data rate loss will be incurred when using the lower-bound compared with the accurate data rate, which is negligible. On the other hand, for the dense scenario, the C-RAN becomes interference limited and the data rate remains fixed for all ranges of the transmit power as expected. It is again observed that the gap between the lower-bound and the exact value is small. Hence, considering the complicated data rate expression in (\ref{finalexpre}) in Appendix \ref{specialcase}, our derived lower-bound expression in (\ref{thirdp}) is much easier to handle and more suitable for algorithm design.

By replacing the data rate $\bar r_k^{(n)}$ in Problems ${\cal P}_1$ and  ${\cal P}_2$ with its lower-bound $\tilde r_k^{(n)}$ given in (\ref{thirdp}) and considering the fact that the minimum rate constraints are met with equality at the optimal point, Problems ${\cal P}_1$ and  ${\cal P}_2$ can be transformed as
\begin{subequations}\label{appstaone}
\begin{align}
{\cal P}_3:\ \mathop {\max }\limits_{{\bf{w}}, {\cal U} \subseteq {\overline {\cal U}} } \quad
& \left| \cal U \right|
\\
\qquad\ \textrm{s.t.}\qquad\!\!\!\!
&{\rm{C3}}, {\rm{C4:}} \  \sum\nolimits_{n \in {\cal N}} {{\tilde  r}_k^{(n)}{({\bf{w}})}}  \ge {R_{k,\min }},\forall k \in {\cal U}, \\
&{\rm{C5:}}\ \sum\nolimits_{k \in {\cal U}_i} {\varepsilon \left( {P_{i,k}^{{\rm{tr}}}({\bf{w}})} \right) {R_{k,\min }}}  \le {C_{i,\max }},\forall i \in {\cal I},
\end{align}
\end{subequations}
and
\begin{subequations}\label{appmainpro}
\begin{align}
{\cal P}_4:\ \mathop {\min }\limits_{{\bf{w}}} \quad
&{{\tilde P}_{{\rm{tot}}}}({\bf{w}}) \triangleq \sum\limits_{i \in {{\cal I}}} {\left\{ {{{{{\eta _i}}}{P_i^{{\rm{tr}}}({\bf{w}})}} \!+\! \varepsilon \left(P_i^{{\rm{tr}}}({\bf{w}}) \right)P_i^{\rm{c}}\!\! + {\rho _i} \sum\limits_{k \in {\cal U}_i^\star} {\varepsilon \left( {P_{i,k}^{{\rm{tr}}}({\bf{w}})} \right)R_{k,\min }}} \right\}} \label{objfunctkklkl}
\\
\qquad\ \textrm{s.t.}\qquad\!\!\!\!
&{\rm{C3}},{\rm{C4}},{\rm{C5}},\nonumber
\end{align}
\end{subequations}
respectively.

In the following, we focus on Problems ${\cal P}_3$ and  ${\cal P}_4$.

\subsection{Problem Analysis}\label{proana}

By adopting the user-centric clustering method in Section II, the number of optimization variables in Problems ${\cal P}_3$ and  ${\cal P}_4$ has been reduced from $NMI\left| {\cal U} \right|$ in fully cooperative transmission scheme to $NM\sum\nolimits_{k \in {\cal U}} {\left| {{{\cal I}_k}} \right|} $ here. By appropriately setting the cluster sizes, the reduced  number of variables $\left( {NM\left( {I\left| {\cal U} \right|{\rm{ - }}\sum\nolimits_{k \in {\cal U}} {\left| {{{\cal I}_k}} \right|} } \right)} \right)$ may be very large, which significantly reduces the computational complexity. In addition, some redundant constraints can be removed, which can additionally reduce the computational complexity. For example, in Fig.~\ref{fig1}, RRH 3 and RRH 5 are not in any UE's candidate serving set, and thus the power constraints associated with RRH 3 and RRH 5  in C3 can be removed. Moreover, if each link supports at most two UEs, then only link 8 (i.e., RRH 8) should be imposed with the fronthaul capacity constraints. Hence, by employing the user-centric clustering with limited cooperation, the computational complexity can be reduced significantly.

However, Problems ${\cal P}_3$ and  ${\cal P}_4$ are still difficult to solve due to the following reasons. Both the objective functions and constraint C5
contain the non-smooth and non-differential indicator function or (and) continuous variables, which are usually named as an MINLP problem. Although the generalized Benders decomposition method \cite{dwkNg2015,Ramamonjison2014} is effective in solving this kind of problems, it is very difficult to directly apply this method to Problems ${\cal P}_3$ and  ${\cal P}_4$ due to the non-convex sum data rate constraints over all multiple SCs. An exhaustive search method can be applied to solve Problems ${\cal P}_3$ and  ${\cal P}_4$. Specifically, to solve Problem ${\cal P}_3$, one should check whether Problem ${\cal P}_3$ is feasible or not for each given user set $\cal U$ and each given set of UE-RRH associations. This requires $O\left( {{2^{\left| {\cal U} \right| + \left| {\cal U} \right|I}}} \right)$ operations, which will become prohibitive for large values of $\left| {\cal U} \right|$ and $I$. In addition, even given the selected UE set ${\cal U}$ and the set of UE-RRH associations, it is still difficult to check the feasibility since constraint C4 is non-convex. Moreover, for dense C-RAN, the complexity associated with the exhaustive search method is unaffordable for BBU pool. Similar difficulties hold for Problem ${\cal P}_4$.

In the next section, we first deal with NPC minimization Problem ${\cal P}_4$ by assuming that the UEs have been selected with feasible beam-vectors, then one low-complexity UE selection algorithm to deal with Problem ${\cal P}_3$ is provided in Section \ref{userselectionalg}.

\section{Low-complexity Algorithm to deal with Problem ${\cal P}_4$}\label{NPCalg}

In this section, we propose a low-complexity algorithm to solve Problem ${\cal P}_4$ when UEs have been selected by using the UE selection algorithms in Section \ref{userselectionalg}, and denote the selected subset of UEs as $\cal U$. As analyzed in Section \ref{proana}, there are  two difficulties to solve Problem ${\cal P}_4$: one is the non-convex sum data rate constraint C4 and the other one is the non-smooth indicator function.

To deal with the first difficulty, we resort to the relationship between the data rate and weighted mean square error (MSE). In \cite{Qingjiang2011}, the authors considered the sum rate maximization problem by showing that maximizing the sum rate is equivalent to minimizing the weighted MSE. Unfortunately, there are two hurdles that preclude the direct application of the technique in \cite{Qingjiang2011}: First, \cite{Qingjiang2011} considered the multiple-antenna UEs with perfect CSI. When each UE has only one antenna with perfect CSI, the rank of channel covariance matrices will be equal to one, i.e., ${\rm{rank}}\left({{\bf{\bar h}}{{_{l,k}^{(n)}}^{\rm{H}}}{\bf{\bar h}}_{l,k}^{(n)}} \right) =1,\forall l,k,n$. However, for the incomplete CSI considered in this paper, the rank of channel covariance matrix may be larger than 1 according to (\ref{akk}), i.e., ${\rm{rank}}\left( {{\bf{A}}_{l,k}^{(n)}} \right) > 1,\forall n,l,k$.
Second, in \cite{Qingjiang2011}, the rate expression is in the objective function, while the rate expressions are in the constraints here.

To resolve the first hurdle, we construct an auxiliary signal transmission model by decomposing each interfering UE into multiple interfering sources. Specifically, for each UE $k$ on SC $n$, since ${{\bf{A}}_{l,k}^{(n)}},\forall l\neq k$, are positive definite matrices, they can be decomposed as
\begin{equation}\label{dhia}
  {\bf{A}}_{l,k}^{(n)} = {\bf{V}}_{l,k}^{(n)}{\bf{V}}_{l,k}^{(n){\rm{H}}},
\end{equation}
where ${\bf{V}}_{l,k}^{(n)} = \left[ {{\bf{v}}_{l,k,1}^{(n)}, \cdots ,{\bf{v}}_{l,k,d_{l,k}^{(n)}}^{(n)}} \right], \forall l\neq k$, with $d_{l,k}^{(n)}$ being the rank of ${{\bf{A}}_{l,k}^{(n)}}$. Then, we construct the following auxiliary signal transmission model for UE $k$
\begin{equation}\label{hdhi}
  {\tilde y}_k^{(n)} = {\bf{\bar h}}_{k,k}^{(n)}{\bf{\bar w}}_k^{(n)}{\tilde s}_k^{(n)} + \sum\nolimits_{l \in {\cal U},l \ne k} {\sum\nolimits_{d = 1}^{d_{l,k}^{(n)}} {{\bf{v}}_{l,k,d}^{(n){\rm{H}}}{\bf{\bar w}}_l^{(n)}{\tilde s}_{l,d}^{(n)}} }  + z_{k}^{(n)},
\end{equation}
where $d_{l,k}^{(n)}$ can be regarded as the number of interfering sources from UE $l$, ${{\bf{v}}_{l,k,d}^{(n){\rm{H}}}}$ can be treated as the CSI from the $d$th interfering source of UE $l$ to UE $k$,  ${\tilde s}_{l,d}^{(n)}$ is the corresponding transmission data. Both ${\tilde s}_{l,d}^{(n)}$ and ${\tilde s}_k^{(n)}$ are assumed to obey the distribution of ${\cal C}{\cal N}({{0}},{{1}})$. The data from different interfering sources are mutually independent and independent of ${\tilde s}_k^{(n)}$. Note that all interfering sources from the same UE use the same beam-vector. By using  the receive decoding $u_k^{(n)}\in \mathbb{C}$ to decode UE $k$'s received signal on SC $n$ \footnote{The decoding parameters $\{u_k^{(n)}\in \mathbb{C}, \forall n,k\}$ can be iteratively calculated at the BBU pool by using the following iterative algorithm. Then BBU pool will send these parameters to the corresponding UEs for decoding their signals. Note that these parameters cannot be included in beam-vectors, otherwise they will affect the transmit power at each RRH.  }, the estimated signal is given by
\begin{equation}\label{decoded}
{{{\hat s}}_k^{(n)}} = {{u}}_k^{(n){\rm{H}}}{{ {\tilde y}}_k^{(n)}},
\end{equation}
 Due to the independence of the transmit data and noise, the mean square error (MSE) matrix at  UE $k$ is given by
 \begin{eqnarray}
 &&{{\epsilon}_k^{(n)}}\left( {{\bf{u}},{\bf{w}}} \right)\nonumber\\
&=&\!\!\!\! {\mathbb{E}_{\{{{\bf{{\tilde s}}},z_k^{(n)}}\}}}\left[ {\left( {{{{{\hat s}}}_k^{(n)}} - {{{\tilde s}}_k^{(n)}}} \right){{\left( {{{{{\hat s}}}_k^{(n)}} - {{{\tilde s}}_k^{(n)}}} \right)}^H}} \right]\nonumber\\
&=&\!\!\!\!\left(\! {u_k^{(n){\rm{H}}}{\bf{\bar h}}_{k,k}^{(n)}{\bf{\bar w}}_k^{(n)} \!-\! 1} \!\right)\!\left( {u_k^{(n){\rm{H}}}{\bf{\bar h}}_{k,k}^{(n)}{\bf{\bar w}}_k^{(n)}\! - \!1} \right)^{\rm{H}}\!\! +\!\! \sum\limits_{l \in {\cal U},l \ne k} \!{{{\left| {u_k^{(n)}} \right|}^2}{{\bf{\bar w}}_l^{(n){\rm{H}}}{\bf{A}}_{l,k}^{(n)}{\bf{\bar w}}_l^{(n)}}} \! + \!\sigma _k^2{\left| {u_k^{(n)}} \right|^2} \label{mse},
\end{eqnarray}
where ${\bf{u}}$ and ${\bf{\tilde s}}$ are the collections of decoding variables and data symbols, respectively, and  (\ref{dhia}) has been used to derive (\ref{mse}).

To deal with the second hurdle, we successfully find a lower bound of the sum rate for each UE and this lower bound is tight at certain point. Then, we replace the sum rate in constraints C4 with its lower bound and iteratively solve the beam-vectors by using the block coordinate decent method. Specifically, defining the following functions:
\begin{equation}\label{hfunction}
\Psi _k^{(n)}\left( {{{\bf{w}}},q_k^{(n)},u_k^{(n)}} \right) = {\log _2}e\left( {\ln( q_k^{(n)}) - q_k^{(n)}\epsilon_k^{(n)}\left( {{\bf{u}},{\bf{w}}} \right) + 1} \right),\forall k \in {\cal U},
\end{equation}
where $q_k^{(n)}\geq 0$ is an introduced variable,  we have the following lemma:

\itshape \textbf{Lemma 1:}  \upshape Given the beam-vectors ${\bf{w}}$, function $\Psi_k^{(n)}\left( {{{\bf{w}}},q_k^{(n)},u_k^{(n)}} \right)$ is a lower bound for $\tilde r_k^{(n)}({\bf{w}})$. In addition, the optimal $u_k^{(n)}$ and  $q_k^{(n)}$ for $\Psi_k^{(n)}\left( {{{\bf{w}}},q_k^{(n)},u_k^{(n)}} \right)$ to achieve $\tilde r_k^{(n)}({\bf{w}})$ are
\begin{eqnarray}
  u_k^{(n)\star} &=& {\left( {{{{\left| {{\bf{\bar h}}_{k,k}^{(n)}{\bf{\bar w}}_k^{(n)}} \right|}^2}} + \sum\nolimits_{l \in {\cal U},l \ne k} {{\bf{\bar w}}{{_l^{(n){\rm{H}}}}}{\bf{A}}_{l,k}^{(n)}} {\bf{\bar w}}_l^{(n)}}+\sigma _k^2 \right)^{ - 1}}{\bf{\bar h}}_{k,k}^{(n)}{\bf{\bar w}}_k^{(n)}, \label{receU} \\
  q_k^{(n)\star} &=& {\left( {\epsilon_k^{(n)}}\left( {{\bf{u}}^{\star},{\bf{w}}} \right) \right)^{ - 1}},\label{receW}
\end{eqnarray}
where $\epsilon_k^{(n)}\left( {{\bf{u}}^{\star},{\bf{w}}} \right)$ is given by
\begin{equation}\label{enk}
 \epsilon_k^{(n)}\left( {{\bf{u}}^{\star},{\bf{w}}} \right) = 1 - \frac{{{{\left| {{\bf{\bar h}}_{k,k}^{(n)}{\bf{\bar w}}_k^{(n)}} \right|}^2}}}{{{{\left| {{\bf{\bar h}}_{k,k}^{(n)}{\bf{\bar w}}_k^{(n)}} \right|}^2} + \sum\limits_{l \in {\cal U},l \ne k} {{\bf{\bar w}}{{_l^{(n)}}^{\rm{H}}}{\bf{A}}_{l,k}^{(n)}} {\bf{\bar w}}_l^{(n)}+\sigma _k^2  }}.
\end{equation}
\itshape \textbf{Proof:}  \upshape
Please see Appendix \ref{prooflemma1}. \hfill $\Box$

By replacing  $\tilde r_k^{(n)}(\bf{w})$ in Problem ${\cal P}_4$ with its lower-bound $\Psi_k^{(n)}\left( {{{\bf{w}}},q_k^{(n)},u_k^{(n)}} \right)$, Problem ${\cal P}_4$ can be transformed into the following optimization problem
\begin{subequations}\label{feasiEquivalentpro}
\begin{align}
{\cal P}_5:\ \mathop {\min }\limits_{{\bf{u},\bf{q},\bf{w}}} \quad
& {{\tilde P}_{{\rm{tot}}}}(\bf{w})\label{vfdaluobj}
\\
\qquad\ \textrm{s.t.}\qquad\!\!\!\!
&\  {{\rm{C3,C5,}}}\nonumber\\
&\ {\rm{C6}}: {\sum\nolimits_{n \in {\cal N}} \Psi_k^{(n)}\left( {{{\bf{w}}},q_k^{(n)},u_k^{(n)}} \right) \ge {R_{k,\min }},\forall k \in {\cal U},}\label{c799}
\end{align}
\end{subequations}
where ${\bf{u}}$ and ${\bf{q}}$ are the collection of variables $\left\{ {u_k^{(n)},\forall n,k} \right\}$ and $\left\{ {q_k^{(n)},\forall n,k} \right\}$, respectively. Note that given  ${\bf{u}}$ and ${\bf{q}}$,  constraint C6 is a convex set over beam-vectors, which is more tractable than Problem ${\cal P}_4$, wherein constraint C4 is  non-convex. Hence, Problem ${\cal P}_5$ can be solved by using the block coordinate decent method: given ${\bf{w}}$, update ${\bf{u}}$ and ${\bf{q}}$ in (\ref{receU}) and (\ref{receW}), respectively; update $\{\alpha_k\}_{k\in { {\cal U}}} $ and ${\bf{w}}$ with fixed ${\bf{u}}$ and ${\bf{q}}$. We only need to deal with the latter one. Given ${\bf{u}}$ and ${\bf{q}}$,  by inserting the MSE expression in (\ref{mse}) into C6, Problem ${\cal P}_5$ can be transformed as
\begin{subequations}\label{wmmseproblem}
\begin{align}
{\cal P}_6:\ \mathop {\min }\limits_{{\bf{w}}} \
& {{\tilde P}_{{\rm{tot}}}}(\bf{w})\label{vfdaluobj}
\\
\ \ \ \textrm{s.t.}\qquad\!\!\!\!
&{{\rm{C3,C5,}}}\nonumber\\
&{\rm{C7}}: \sum\limits_{n \in {\cal N}} {q_k^{(n)}\left( {{{\left| {u_k^{(n)}} \right|}^2}{\bf{\bar w}}_k^{(n){\rm{H}}}{\bf{\bar h}}_{k,k}^{(n){\rm{H}}}{\bf{\bar h}}_{k,k}^{(n)}{\bf{\bar w}}_k^{(n)} - u_k^{(n){\rm{H}}}{\bf{\bar h}}_{k,k}^{(n)}{\bf{\bar w}}_k^{(n)} - u_k^{(n)}{\bf{\bar w}}_k^{(n){\rm{H}}}{\bf{\bar h}}_{k,k}^{(n){\rm{H}}}} \right)}  \nonumber\\
 &\ + \sum\limits_{n \in {\cal N}} {\sum\limits_{l \in {\cal U},l \ne k} {q_k^{(n)}{{\left| {u_k^{(n)}} \right|}^2}{\bf{\bar w}}{{_l^{(n){\rm{H}}}}}{\bf{A}}_{l,k}^{(n)}} {\bf{\bar w}}_l^{(n)}}   \le {\omega _k},\forall k \in {\cal U},\label{c8dfd}
\end{align}
\end{subequations}
where  ${\omega_k} = \sum\nolimits_{n \in {\cal N}} {\left[ {\ln \left( {q_k^{(n)}} \right) - q_k^{(n)}\sigma _k^2{{\left| {u_k^{(n)}} \right|}^2}} -q_k^{(n)}\right]}  + N-{R_{k,\min }}\ln 2$.

Now, we deal with the second difficulty: the non-smooth indicator function $\varepsilon \left( \cdot \right)$ in (\ref{indifunc}) in the objective function and C5 in Problem ${\cal P}_6$. The non-smooth indicator function is approximated  as a fractional function ${f_\theta }(x) = \frac{x}{{x + \theta }}$, where $\theta$ is a very small positive value that controls the smoothness of approximation\footnote{Smaller value of $\theta$ will result in more accurate approximation but leads to less smoothness in the function, while larger value of $\theta$ leads to high approximation error. From simulations, we find that $\theta=10^{-5}$ can achieve a good balance between smoothness and approximation accuracy. In the simulations, when transmit power for each link is smaller than $10^{-8}$ watt, the transmit power is set to be zero. The effect on the rate of each user can be negligible. In addition, for practical analog to digital conversion (ADC) or digital to analog conversion (DAC), there is a minimum required power to activate it. Hence, when the transmit power is very small, it can be ignored.  }.
 Then, ${{\tilde P}_{{\rm{tot}}}}(\bf{w})$ can be approximated as
\begin{eqnarray}
  {{\tilde P}_{{\rm{tot}}}}(\bf{w}) &\approx& \sum\nolimits_{i \in {\cal I}} {\left\{ {{{{{\eta _i}}}{P_i^{{\rm{tr}}}(\bf{w})}} + {f_\theta }\left( {P_i^{{\rm{tr}}}(\bf{w})} \right)P_i^{\rm{c}} + {\rho _i}\sum\nolimits_{k \in {{\cal U}_i}} { {f_\theta }\left( {P_{i,k}^{{\rm{tr}}}(\bf{w})}  \right) {R_{k,\min }}} } \right\}}  \\
  \!\!\! &\buildrel \Delta \over =&\!\!\! {{\hat P}_{{\rm{tot,}}\theta }}(\bf{w}).
\end{eqnarray}
Note that for any positive $\theta$, the fractional function ${f_\theta }(x)$ is strictly smaller than one. Hence, ${{\hat P}_{{\rm{tot,}}\theta }}(\bf{w})$ is actually the lower bound of ${{\tilde P}_{{\rm{tot}}}}(\bf{w})$. However, this gap is negligible when $\theta$ is very small and $x$ is comparatively large. By replacing the indicator function in  Problem ${\cal P}_6$ with $f_{\theta}(x)$,  we have
\begin{subequations}\label{apprix}
\begin{align}
{\cal P}_7:\ \mathop {\min }\limits_{ {\bf{w}}} \quad
& {{\hat P}_{{\rm{tot,}}\theta }}(\bf{w}) \label{objcedj}
\\
\qquad\ \textrm{s.t.}\qquad\!\!\!\!
&\  {{\rm{C3,C7,}}}\nonumber\\
&\ {\rm{C8}}: \sum\nolimits_{k \in {\cal U}_i} {{f_\theta }\left( {P_{i,k}^{{\rm{tr}}}(\bf{w})}  \right) {R_{k,\min }}}  \le {C_{i,\max }},\forall i \in {\cal I}.
\end{align}
\end{subequations}
Problem ${\cal P}_7$ is much more tractable than Problem ${\cal P}_6$ since both the objective function and constraints in Problem ${\cal P}_7$ are differentiable and continuous. Although Problem ${\cal P}_7$  is still nonconvex due to the concavity  of ${f_\theta }\left(\cdot \right)$, it is a well-known difference of convex (d.c.) program, which can be efficiently solved by the SCA method \cite{dinh2010local}. The main idea of this method is to approximate the concave function as its first order Taylor expansion. Specifically, by using the concavity of ${f_\theta }(\cdot)$, one has
\begin{equation}\label{first}
{f_\theta }\left( {P_i^{{\rm{tr}}}({\bf{w}})} \right) \le {f_\theta }\left( {P_i^{{\rm{tr}}}({\bf{w}}(t))} \right) + {{ \beta }_i}(t)\left( {P_i^{{\rm{tr}}}({\bf{w}}) - P_i^{{\rm{tr}}}({\bf{ w}}(t))} \right),
\end{equation}
\begin{equation}\label{second}
 {f_\theta }\left( {P_{i,k}^{{\rm{tr}}}({\bf{w}})} \right) \le {f_\theta }\left( {P_{i,k}^{{\rm{tr}}}({\bf{ w}}(t))} \right) + {{ \chi  }_{i,k}}(t)\left( {P_{i,k}^{{\rm{tr}}}({\bf{w}}) - P_{i,k}^{{\rm{tr}}}({\bf{ w}}(t))} \right)
\end{equation}
where ${{\bf{ w}}(t)}$ is a collection of beam-vectors at the $t^{\rm{th}}$ iteration, ${{ \beta }_i}(t)$ and ${{  \chi }_{i,k}}(t)$ are given by
\begin{equation}\label{betagama}
{{ \beta }_i}(t) = {{f}_\theta^{'} }\left( {P_i^{{\rm{tr}}}({\bf{w}}(t))} \right),{{  \chi }_{i,k}}(t) = {{f}_\theta^{'} }\left( {P_{i,k}^{{\rm{tr}}}({\bf{w}}(t))} \right),
\end{equation}
where ${{f}_\theta^{'} }\left( x \right)$ denotes the first-order derivative of $x$. By replacing ${f_\theta }\left( {P_i^{{\rm{tr}}}({\bf{w}})} \right)$ and ${f_\theta }\left( {P_{i,k}^{{\rm{tr}}}({\bf{w}})} \right)$ in Problem ${\cal P}_7$ with the right hand side (RHS) of (\ref{first}) and (\ref{second}), respectively, one can solve the following optimization problem in the $(t+1)^{\rm{th}}$ iteration
\begin{subequations}\label{finaloptim}
\begin{align}
{\cal P}_8:\ \mathop {\min }\limits_{{\bf{w}}} \quad
& \sum\limits_{n \in {\cal N}} {\sum\limits_{k \in {\cal U}} {{\bf{\bar w}}_k^{(n){\rm{H}}}{\bf{G}}_k(t){\bf{\bar w}}_k^{(n)}} }  \label{objfdcedj}
\\
\qquad\ \textrm{s.t.}\qquad\!\!\!\!
&\  {{\rm{C3,C7,}}}\nonumber\\
&\ {\rm{C9}}: \sum\limits_{n \in {\cal N}} {\sum\limits_{k \in {{\cal U}_i}} {{\tau _{i,k}}(t){{\left\| {{\bf{w}}_{i,k}^{(n)}} \right\|}^2}} }  \le {{\tilde C}_{i,\max }}(t), \forall i\in {\cal I}.\label{c10dsh}
\end{align}
\end{subequations}
where ${\bf{G}}_k(t)$ is given by ${\bf{G}}_k(t) = {\rm{blkdiag}}\left( {{\kappa _{s_1^k}}(t){{\bf{I}}_{M \times M}}, \cdots ,{\kappa _{s_{\left| {{{\cal I}_k}} \right|}^k}}(t){{\bf{I}}_{M \times M}}} \right)$ with ${\kappa _{s_i^k}}(t) = \left( {{\eta _{s_i^k}} + {\beta _{s_i^k}}(t)P_{s_i^k}^{\rm{c}} + {\rho _{s_i^k}}{ \chi_{s_i^k,k}}(t){R_{k,\min }}} \right),i = 1, \cdots ,\left| {{{\cal I}_k}} \right|$, ${\tau _{i,k}}(t) = { \chi _{i,k}}(t){R_{k,\min }}$, and ${{\tilde C}_{i,\max }}(t) = {C_{i,\max }} - \sum\nolimits_{k \in {{\cal U}_i}} {\left( {{f_\theta }\left( {P_{i,k}^{{\rm{tr}}}({\bf{ w}}(t))} \right) - { \chi _{i,k}}(t)P_i^{{\rm{tr}}}({\bf{w}}(t))} \right){R_{k,\min }} }$.
Note that some constant terms in the RHS of (\ref{first}) and (\ref{second}) are omitted in (\ref{objfdcedj}). Obviously, ${{{\bf{G}}_k{(t)}}}$ is a positive definite matrix and all constraints form a convex set. Then Problem ${\cal P}_8$ is a convex problem. The details to solve it will be given in the next subsection.

Based on the above analysis, an iterative algorithm is given to solve Problem ${\cal P}_4$. A straightforward way to solve Problem ${\cal P}_4$ would involve two layers: the inner layer to solve Problem ${\cal P}_8$ by using the SCA method given ${\bf{u}}$ and ${\bf{q}}$; the outer layer to update ${\bf{u}}$ and ${\bf{q}}$ by using (\ref{receU}) and (\ref{receW})  given ${\bf{w}}$. Although the inner layer is guaranteed to converge to a Karush-Kuhn-Tucker (KKT) point of Problem ${\cal P}_7$ as proved in \cite{Panwcl}, this two-layer algorithm will incur high computational complexity. Instead, we merge these two layers into one layer and update $\{\beta _i{(t)}, {{\tilde C}_{i,\max }}(t), \forall i\}$, $\{ \chi _{i,k}{(t)}, {\tau _{i,k}}(t), \forall i,k\}$, ${\bf{u}}(t)$ and ${\bf{q}}(t)$  at the same layer, as given in Algorithm \ref{algorithmiter}. Fortunately, Algorithm \ref{algorithmiter} is guaranteed to converge, as proved in Theorem 1.

\begin{algorithm}
\caption{Iterative Algorithm to Solve Problem ${\cal P}_4$}\label{algorithmiter}
\begin{algorithmic}[1]
\STATE Initialize  the iterative number $t=1$,  error tolerance $\delta $. Initialize ${\bf{w}}{(0)}$ with the output from the UE selection algorithm in Section \ref{userselectionalg}, calculate $\{\beta _i{(0)},  \chi _{i,k}{(0)},  {\bf{G}}_k{(0)}, {\tau _{i,k}}(0), {{\tilde C}_{i,\max }}(0),  \forall i,k \}$, calculate ${\bf{u}}{(0)}$ and ${\bf{q}}{(0)}$ by using (\ref{receU}) and (\ref{receW}) with ${\bf{w}}{(0)}$,  calculate the objective value of Problem ${\cal P}_7$, denoted as ${\rm{Obj(}}{{\bf{w}}{(0)}}{\rm{)}}$.
 \STATE Solve Problem ${\cal P}_8$ to get ${\bf{w}}{(t)}$ with $\{\beta _i{(t-1)},  \chi _{i,k}{(t-1)},  {\bf{G}}_k{(t-1)}, {\tau _{i,k}}(t-1), {{\tilde C}_{i,\max }}(t-1), \forall i,k \}$, ${\bf{u}}{(t-1)}$ and ${\bf{q}}{(t-1)}$;
 \STATE Update $\{\beta _i{(t)}, \chi _{i,k}{(t)},  {\bf{G}}_k{(t)},  {\tau _{i,k}}(t), {{\tilde C}_{i,\max }}(t), \forall i,k \}$ with ${\bf{w}}{(t)}$;
 \STATE Update ${\bf{u}}{(t)}$ and ${\bf{q}}{(t)}$ by using (\ref{receU}) and (\ref{receW}) with  ${\bf{w}}{(t)}$;
 \STATE If  ${{\left| {{\rm{Obj(}}{{\bf{w}}{(t - 1)}}{\rm{) - Obj(}}{{\bf{w}}{(t)}}{\rm{)}}} \right|} \mathord{\left/
 {\vphantom {{\left| {{\rm{Obj(}}{{\bf{V}}^{(n - 1)}}{\rm{) - Obj(}}{{\bf{V}}^{(n)}}{\rm{)}}} \right|} {{\rm{Obj(}}{{\bf{w}}^{(t)}}{\rm{)}}}}} \right.
 \kern-\nulldelimiterspace} {{\rm{Obj(}}{{\bf{w}}{(t)}}{\rm{)}}}} < \delta  $, terminate.  Otherwise, set $t \leftarrow t + 1$  and go to step 2.
\end{algorithmic}
\end{algorithm}

\itshape \textbf{Theorem 1:}  \upshape Given the feasible initial input ${\bf{w}}(0)$, Algorithm \ref{algorithmiter} is guaranteed to converge both in objective value and variables.

\itshape \textbf{Proof:}  \upshape Please see Appendix \ref{prooftheorem1}. \hfill $\Box$

\subsection{Lagrange dual decomposition method to solve Problem ${\cal P}_8$}\label{lowcom}

In Step 2 of Algorithm \ref{algorithmiter}, Problem ${\cal P}_8$ should be solved. Since both the maximum power limit and fronthaul capacity limit are positive, i.e., ${C_{i,\max }} > 0,{P_{i,\max }} > 0,\forall i \in {\cal I}$, the Slater's condition of Problem ${\cal P}_8$ is satisfied and the duality gap between Problem ${\cal P}_8$ and its dual problem is zero \cite{boyd2004convex}. Hence, Problem ${\cal P}_8$ can be equivalently solved by solving its dual problem. In the following, we derive the optimal form of ${\bf{w}}$  for Problem ${\cal P}_8$ by using the Lagrange dual decomposition method. For notation simplicity, we omit the iteration index $t$ in the following derivations.

Define the following block diagonal matrices
\begin{equation}\label{blocmat}
  {{\bf{B}}_{i,k}}{\rm{ = diag}}\left\{ {\overbrace {{{\bf{0}}_{1 \times M}}}^{s_1^k}, \cdots ,\overbrace {{{\bf{1}}_{1 \times M}}}^{s_j^k},\overbrace {{{\bf{0}}_{1 \times M}}}^{s_{j + 1}^k}, \cdots ,\overbrace {{{\bf{0}}_{1 \times M}}}^{s_{\left| {{{\cal I}_{ k}}} \right|}^k}} \right\},\ {\rm{if }}\ s_j^k = i,\forall i \in {\cal I},k \in {\cal U},
\end{equation}
then ${\left\| {{\bf{w}}_{i,k}^{(n)}} \right\|^2} = {\bf{\bar w}}_k^{(n){\rm{H}}}{{\bf{B}}_{i,k}}{\bf{\bar w}}_k^{(n)}$. With some manipulations, the Lagrangian function of Problem ${\cal P}_8$ is given by
\begin{eqnarray}
&&{\cal L}\left( {{\bf{w}},{\bm{\lambda}} ,{\bm{\mu}} ,{\bm{\nu}} } \right)\nonumber\\
 &=& \sum\limits_{n \in {\cal N}} {\sum\limits_{k \in {\cal U}} {{\bf{\bar w}}_k^{(n){\rm{H}}}{\bf{J}}_k^{(n)}{\bf{\bar w}}_k^{(n)}} }  - \sum\limits_{n \in {\cal N}} {\sum\limits_{k \in {\cal U}} {{\nu _k}q_k^{(n)}\left( {u_k^{(n){\rm{H}}}{\bf{\bar h}}_{k,k}^{(n)}{\bf{\bar w}}_k^{(n)} + u_k^{(n)}{\bf{\bar w}}_k^{(n){\rm{H}}}{\bf{\bar h}}_{k,k}^{(n){\rm{H}}}} \right)} }\nonumber \\
&& +\ln 2 \sum\limits_{k \in {\cal U}} {{\nu _k}{R_{k,\min }}}  - \sum\limits_{i \in {\cal I}} {{\lambda _i}{P_{i,\max }}}  - \sum\limits_{i \in {\cal I}} {{\mu _i}} {{\tilde C}_{i,\max }} - \sum\limits_{k \in {\cal U}} {{\nu _k}{\omega _k}},
\end{eqnarray}
where ${\bm{\lambda}} ,{\bm{\mu}} ,{\bm{\nu}}$ are the collections of non-negative Lagrangian multipliers corresponding to ${\rm{C3}}$ in (\ref{powercons}), ${\rm{C9}}$ in (\ref{c10dsh}) and $\rm{C7}$ in (\ref{c8dfd}), respectively, ${\bf{J}}_k^{(n)}$ is given by
\begin{equation}\label{jkk}
  {\bf{J}}_k^{(n)} = {{\bf{G}}_k} + \sum\limits_{i \in {{\cal I}_k}} {\left( {{\lambda _i} + {\mu _i}{\tau _{i,k}}} \right){{\bf{B}}_{i,k}}}  + {\nu _k}q_k^{(n)}{\left| {u_k^{(n)}} \right|^2}{\bf{\bar h}}_{k,k}^{(n){\rm{H}}}{\bf{\bar h}}_{k,k}^{(n)} + \sum\limits_{l \in {\cal U},l \ne k} {{\nu _l}q_l^{(n)}{{\left| {u_l^{(n)}} \right|}^2}{\bf{A}}_{k,l}^{(n)}}.
\end{equation}

Then, the dual function is given by
\begin{eqnarray}
&&g\left( {{\bm{\lambda}} ,{\bm{\mu}} ,{\bm{\nu}}} \right)\\
\!\!\! &=&\!\!\! \mathop {\min }\limits_{{\bf{w}}} {\cal L}\left( {{\bf{w}},{\bm{\lambda}} ,{\bm{\mu}} ,{\bm{\nu}} } \right)\\
\!\!\!&=&\!\!\! \mathop {\min }\limits_{{\bf{w}}} \sum\limits_{n \in {\cal N}} {\sum\limits_{k \in {\cal U}} {{\bf{\bar w}}_k^{(n){\rm{H}}}{\bf{J}}_k^{(n)}{\bf{\bar w}}_k^{(n)}} }  - \sum\limits_{n \in {\cal N}} {\sum\limits_{k \in {\cal U}} {{\nu _k}q_k^{(n)}\left( {u_k^{(n){\rm{H}}}{\bf{\bar h}}_{k,k}^{(n)}{\bf{\bar w}}_k^{(n)} + u_k^{(n)}{\bf{\bar w}}_k^{(n){\rm{H}}}{\bf{\bar h}}_{k,k}^{(n){\rm{H}}}} \right)} } \nonumber\\
 \!&&  +\ln 2 \sum\limits_{k \in {\cal U}} {{\nu _k}{R_{k,\min }}}  - \sum\limits_{i \in {\cal I}} {{\lambda _i}{P_{i,\max }}}  - \sum\limits_{i \in {\cal I}} {{\mu _i}} {{\tilde C}_{i,\max }} - \sum\limits_{k \in {\cal U}} {{\nu _k}{\omega _k}}.\label{dualpro}
\end{eqnarray}
Obviously, Problem (\ref{dualpro}) is a strictly convex problem and the optimal solution can be easily obtained from its first-order optimality condition as:
\begin{equation}\label{Vk}
  {\bf{\bar w}}_k^{(n)\star} = {\nu _k}q_k^{(n)}u_k^{(n)}{\left( {{\bf{J}}_k^{(n)}} \right)^{ - 1}}{\bf{\bar h}}_{k,k}^{(n){\rm{H}}},\forall n \in {\cal N},k \in {\cal U}.
\end{equation}
By inserting the solution of  $\{{\bf{w}}_k^{\star}, k\in { {\cal U}}\} $ in (\ref{Vk}) into (\ref{dualpro}), the dual function can be rewritten as
\begin{eqnarray}
g({\bm{\lambda}} ,{\bm{\mu}} ,{\bm{\nu}}) &=&   - \sum\limits_{n \in {\cal N}} {\sum\limits_{k \in {\cal U}} {\nu _k^2q_k^{(n)2}{{\left| {u_k^{(n)}} \right|}^2}{\bf{\bar h}}_{k,k}^{(n)}{{\left( {{\bf{J}}_k^{(n)}} \right)}^{ - 1}}{\bf{\bar h}}_{k,k}^{(n){\rm{H}}}} } \nonumber\\
 && +\ln 2 \sum\limits_{k \in {\cal U}} {{\nu _k}{R_{k,\min }}}  - \sum\limits_{i \in {\cal I}} {{\lambda _i}{P_{i,\max }}}  - \sum\limits_{i \in {\cal I}} {{\mu _i}} {{\tilde C}_{i,\max }} - \sum\limits_{k \in {\cal U}} {{\nu _k}{\omega _k}} .\label{dualfunc}
\end{eqnarray}
Hence,  the dual problem of Problem ${\cal P}_9$ is given by
\begin{equation}\label{dualproblem}
  \mathop {\max }\limits_{\left\{ {{\lambda _i} \ge 0,{\mu _i} \ge 0,{\nu _k} \ge 0,\forall k,i} \right\}} g({\bm{\lambda}} ,{\bm{\mu}} ,{\bm{\nu}} ).
\end{equation}
Fortunately, the objective function of the dual Problem (\ref{dualproblem}) is differentiable and dual problem is  a convex optimization problem as defined in \cite{boyd2004convex}. Hence, the classic descent methods such as the gradient descent method can be applied to solve it as detailed in \cite{boyd2004convex}.

\itshape \textbf{Remark 1 - Parallel Computations:}  \upshape  Note that for given Lagrangian multipliers, the optimal beam-vectors $\{ {\bf{\bar w}}_k^{(n)\star}, \forall k\}$ can be obtained in  (\ref{Vk}) in closed forms for each SC in parallel. In C-RAN, multiple-core processors or multiple virtual machines (VMs) are aggregated together in the BBU pool, which entails C-RAN to be capable of the parallel computation. Hence, the Lagrange dual decomposition method can run smoothly under the C-RAN architecture.

\section{A Low-complexity UE Selection Algorithm}\label{userselectionalg}

In this section, we propose a low-complexity UE selection algorithms to deal with Problem ${\cal P}_3$: the bisection UE selection  algorithm, the complexity of which increases logarithmically with the number of UEs $K$.

Inspired by the UE selection problem formulations (28)-(30) in \cite{Matskani2008}, we first construct the following alternative optimization problem by introducing a series of auxiliary variables $\{\varphi_k\}_{k\in {\bar {\cal U}}} $\footnote{The authors in \cite{Matskani2008} considered the single-channel case by introducing auxiliary variables $\{s_k,\forall k\}$ in each UE's useful signal power. When all the optimal $\{s_k,\forall k\}$ are no larger than zeros, all UEs can be admitted. The method in \cite{Matskani2008} cannot be directly extended to our work since we consider the multi-channel case. Instead, we introduce the auxiliary variables $\{\varphi_k\}_{k\in {\bar {\cal U}}} $ on the right hand side of constraint C10. When all the optimal  $\{\varphi_k\}_{k\in {\bar {\cal U}}} $ are equal to one, all UEs can be admitted. Note that  \cite{Matskani2008} optimized the UE selection and transmit power minimization problem simultaneously by introducing a large $M$. In our paper, we consider each problem individually. The reason is that we find from simulations that the big $M$ is difficult to choose and unproperly chosen $M$ may result in unexpected results.}:
\begin{subequations}\label{feasibilityproblem}
\begin{align}
{\cal P}_9:\ \mathop {\min }\limits_{\{\varphi_k\}_{k\in {\bar{\cal U}}}, \bf{w}} \quad
& \sum\nolimits_{k\in {{\cal U}}} {{{({\varphi _k} - 1)}^2}}  \label{valuobjfref}
\\
\qquad\ \textrm{s.t.}\qquad\!\!\!\!
&\  {{\rm{C3,C5,}}}\nonumber\\
&\ {\rm{C10}}: {\sum\nolimits_{n \in {\cal N}} {\tilde r_k^{(n)}(\bf{w})}  \ge \varphi_k^2{R_{k,\min }},\forall k \in \bar{\cal U},}\label{c66}
\end{align}
\end{subequations}
Obviously, Problem ${\cal P}_9$ is always feasible since at least $\{\varphi_k=0, {{\bf{w}}_k^{(n)}} = {\bf{0}},\forall k \in { \bar{\cal U}}, n\in {\cal N}\}$ is a feasible solution. In addition, it is easy to verify that the optimal $\{\varphi_k, k\in {\bar{\cal U}} \}$ should lie between zero and one, i.e.,  $0 \le {\varphi _k} \le 1,\forall k \in  \bar{\cal U} $. If UE $k$ can be admitted, the optimal $\varphi_k$ must be equal to one. This can be easily proved by contradiction. Denote the solution  of $\{\varphi_k\}_{k\in {\bar{\cal U}}}$ as ${\{ \varphi _k^\star\} _{k \in \bar{\cal U} }}$. If  $\varphi _k^\star = 1, \forall k\in { \bar{\cal U}}$, all UEs can be admitted in the network and output the corresponding optimal beam-vectors for the initial solution for Algorithm \ref{algorithmiter} in Section \ref{NPCalg}. Otherwise, some UEs should be removed. Intuitively, the UE with a smaller $\varphi _k^\star$ should have a higher priority to be removed since it has the largest gap away from its rate targets. Hence, we sort ${\{ \varphi _k^\star\} _{k \in  \bar{\cal U} }}$ in the ascending order: $\varphi _{{\pi _1}}^\star \le  \cdots  \le \varphi _{{\pi _K}}^\star$. Then admitting the maximum number of UEs is equivalent to finding a minimum ${L_0}$ such that all the users in $ {\cal U}  = \left\{ {{\pi _{{L_0} + 1}}, \cdots ,{\pi _K}} \right\}$ can be supported by C-RAN with ${L_0} = 1, \cdots ,K - 1$. The bisection search procedure can be adopted to determine the minimum ${L_0}$. In each iteration of the bisection UE search algorithm, we only need to check whether the C-RAN can support all users in $ {\cal U}$ or not. Hence, in each iteration, we  need to solve the following optimization problem
\begin{subequations}\label{bisectionproblem}
\begin{align}
{\cal P}_{10}:\ \mathop {\min }\limits_{\varphi, \bf{w}} \quad
& {({\varphi } - 1)}^2 \label{valuobj}
\\
\qquad\ \textrm{s.t.}\qquad\!\!\!\!
&\  {{\rm{C3,C5,}}}\nonumber\\
&\ {\rm{C11}}: {\sum\nolimits_{n \in {\cal N}} {\tilde r_k^{(n)}(\bf{w})}  \ge \varphi^2{R_{k,\min }},\forall k \in {\cal U},}
\end{align}
\end{subequations}
where  $\varphi$ is the introduced optimization variable. Obviously, when the optimal $\varphi^\star$ is equal to one, Problem ${\cal P}_{10}$ is feasible. Problem ${\cal P}_{10}$ can be similarly solved by using Algorithm \ref{algorithmiter}. Note that all UEs' rate requirements in ${\cal P}_{10}$ use the same $\varphi $ and thus Problem ${\cal P}_{10}$ has less variables than Problem ${\cal P}_{9}$.

Finally, the bisection search method is summarized in Algorithm \ref{bisecselece}. Notice that Problem ${\cal P}_{10}$ only needs to be solved  no more than $\left\lceil {{{\log }_2}(1 + K)} \right\rceil $ times.

\begin{algorithm}
\caption{Bisection UE Selection (BUES) Algorithm}\label{bisecselece}
\begin{algorithmic}[1]
\STATE Solve Problem ${\cal P}_{9}$.
\begin{enumerate}
  \item If $\varphi _k^\star = 1, \forall k\in { \bar{\cal U}}$, terminate and all UEs can be supported, output the corresponding optimal beam-vectors for the initial point for Algorithm \ref{algorithmiter};
  \item If there exists at least one UE $k$ such that $\varphi _k^\star <1$, sort all ${\{ \varphi _k^\star\} _{k \in {\cal U}}}$ in the ascending order: $\varphi _{{\pi _1}}^\star \le  \cdots  \le \varphi _{{\pi _K}}^\star$, go to step 2;
\end{enumerate}
\STATE Set ${ {\cal U}} = \{{\pi _K}\}$, solve Problem ${\cal P}_{10}$:
\begin{enumerate}
  \item If $\varphi _{\pi _K}^\star = 1$, go to step 3;
  \item Otherwise, terminate and claim that no UE can be supported;
\end{enumerate}
\STATE Initialize ${L_{{\rm{low}}}} = 0,{L_{{\rm{up}}}} = K$;
\STATE Repeat
\begin{enumerate}
  \item Set $l \leftarrow \left\lfloor {\frac{{{L_{{\rm{low}}}} + {L_{{\rm{up}}}}}}{2}} \right\rfloor $;
  \item Solve Problem ${\cal P}_{10}$ with ${\cal U}  = \left\{ {{\pi _{l + 1}}, \cdots ,{\pi _K}} \right\}$. If $\varphi^\star = 1$, set ${L_{{\rm{up}}}} = l$; Otherwise, set ${L_{{\rm{low}}}} = l$;
\end{enumerate}
 \STATE Until ${L_{{\rm{up}}}} - {L_{{\rm{low}}}} = 1$. Output the optimal active UE set  ${\cal U}  = \left\{ {{\pi _{{L_{\rm{low}}} + 1}}, \cdots ,{\pi _K}} \right\}$ and the corresponding optimal beam-vectors.
\end{algorithmic}
\end{algorithm}

\section{Simulation Results}\label{simlresult}

\subsection{System parameters}
In this section, we present simulation results to evaluate the performance of the proposed algorithms. The dense C-RAN is within a square area of coordinates $[-1000, 1000]\times [-1000, 1000]$ meters. Both the UEs and RRHs are assumed to be independently and uniformly distributed in this square area. The channel model consists of three parts: 1) the channel path-loss modeled as $PL_{i,k} = 148.1 + 37.6{\log _{10}}d_{i,k}\ ({\rm{dB}})$ \cite{access2010further}, where $d_{i,k}$ (in km) is the distance between the $i$th RRH to the $k$th UE; 2) the log-normal shadowing with zero mean and 8 dB standard derivation; 3) small-scale Rayleigh fading with zero mean and unit variance. All the UEs are assumed to have the same rate requirements, i.e., $R_{k,\rm{min}}=R_{\rm{min}}, \forall k$. For ease of exposition, each fronthaul link is assumed to have the same capacity constraints, i.e., $C_{i,\max}=C_{\max}, \forall i$. Then, normalized fronthaul capacity is considered, i.e., ${\tilde C_{\max }} = {{{C_{\max }}} \mathord{\left/
 {\vphantom {{{C_{\max }}} R}} \right.
 \kern-\nulldelimiterspace} R_{\rm{min}}}$, which represents the maximum number of UEs that can be supported on each fronthaul link is the same. It is assumed that each UE is potentially served by its nearest $X$ RRHs, i.e., $\left| {{{\cal I}_k}} \right| = X,\forall k$. Also, each UE is assumed to measure its channel vectors to its nearest $Y$ RRHs, i.e., $\left| {{\widetilde{{\cal I}}_k}} \right| = Y,\forall k$. Unless stated otherwise, the system parameters are set as follows: $M=2$, $K=16$, $I=20$, $N=3$,  system bandwidth $B=10\  \rm{MHz}$, error tolerance $\delta=10^{-3}$,  noise power spectral density is -174 dBm/Hz, $P_i^{{\rm{active}}}=6.8\  {\rm{Watt}}, P_i^{{\rm{sleep}}}=4.3\  {\rm{Watt}}, \eta _i=4, \rho_i=0.5, P_{i,\rm{max}}=2\  {\rm{Watt}}, \forall i$, $\theta=10^{-5}$, $R_{\rm{min}}=15\ \rm{bit/s/Hz}$, ${\tilde C_{\max }} =3$, $X = 3$, $Y=6$.
\subsection{Numerical Results}
\begin{figure}
\begin{minipage}[t]{0.485\linewidth}
\centering
\includegraphics[width=2.8in]{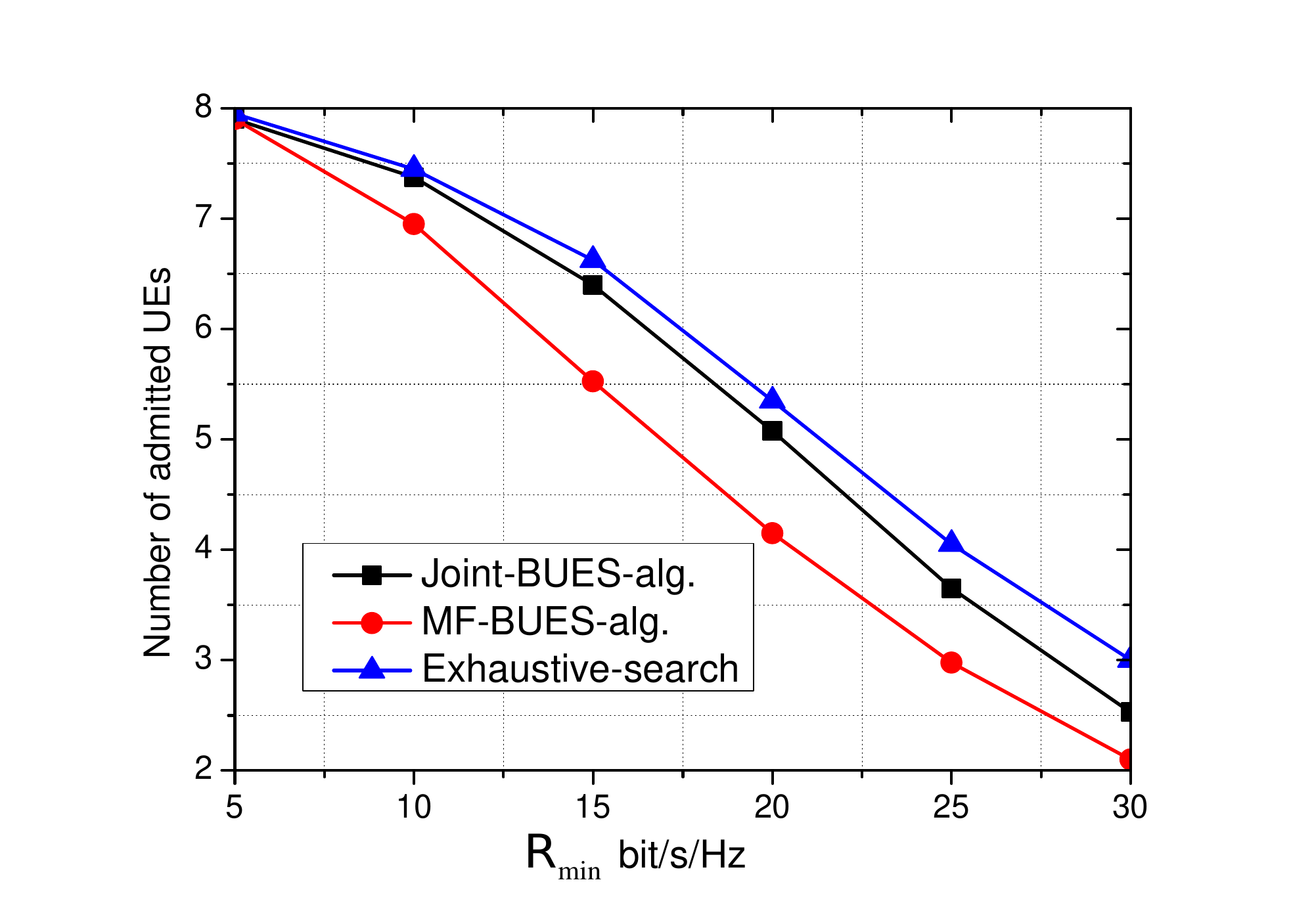}
\caption{The average number of admitted UEs for different algorithms with $K=8$, $I=12$, $P_{i,\rm{max}}=2\  {\rm{Watt}}, \forall i$, $X = 3$, and $Y=6$.}
\label{fig2}
\end{minipage}%
\hfill
\begin{minipage}[t]{0.485\linewidth}
\centering
\includegraphics[width=2.8in]{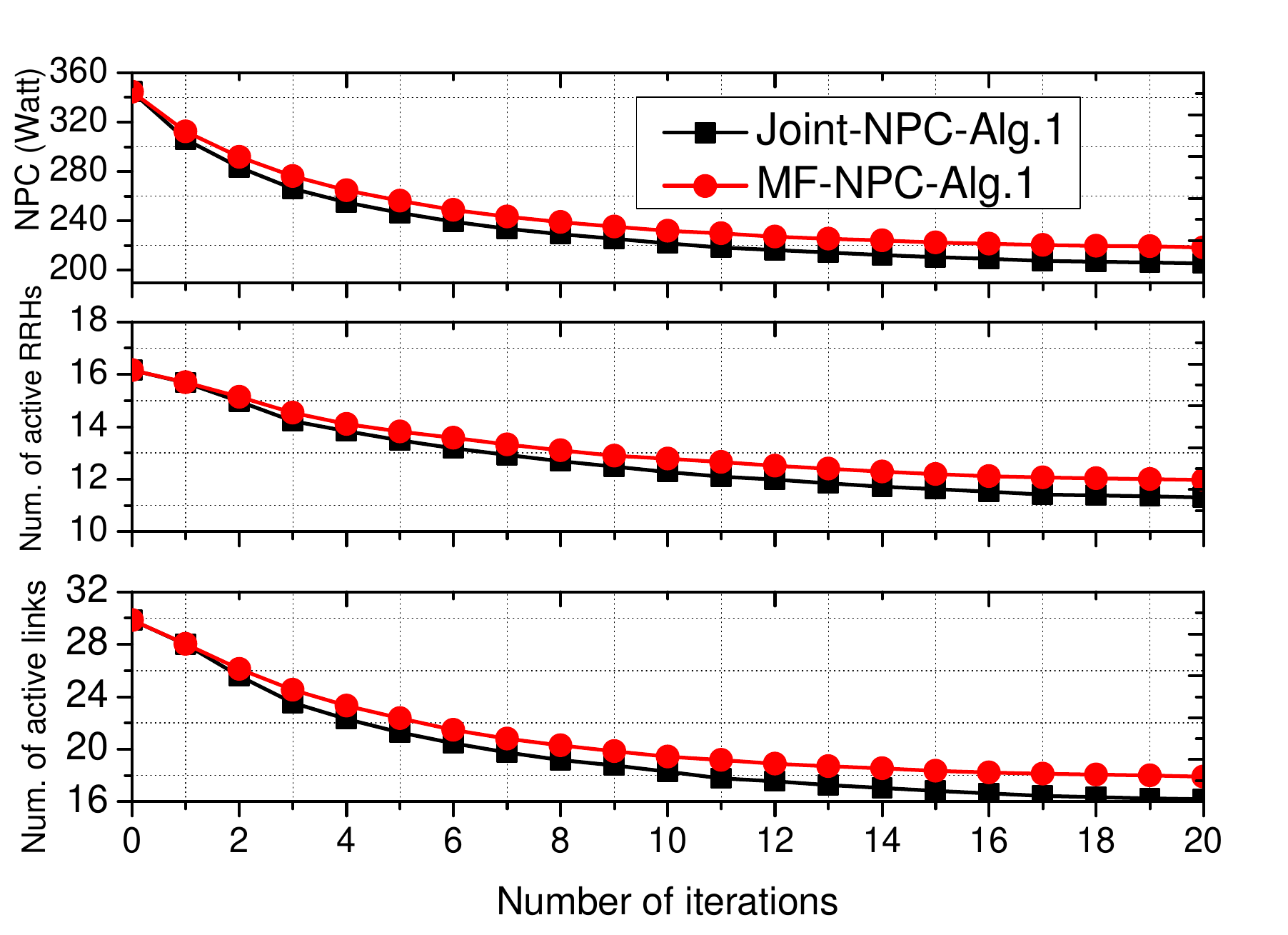}
\caption{Convergence behaviour for the NPC minimization algorithm.}
\label{fig3}
\end{minipage}
\end{figure}

\subsubsection{Performance of the UE selection algorithm}

Fig.~\ref{fig2} shows the average number of admitted UEs for three different algorithms. Specifically, `Joint-BUES-alg.' denotes the joint beam direction and power allocation optimization algorithm in  Algorithm \ref{bisecselece}, while  `MF-BUES-alg.' represents that the beam directions are fixed to be the channel direction, and the power allocation problem is solved by using Algorithm \ref{bisecselece}. Note that beam direction is not optimized in  `MF-BUES-alg.'. This scheme has lower complexity than `Joint-BUES-alg.', but incurs inferior performance as seen in the following examples. For comparison, the optimal performance obtained by exhaustive search \footnote{There are many existing MINLP solvers to solve the MINLP problems, such as the generalized Benders decomposition method in \cite{Ramamonjison2014,DWK16TWC} and the branch-and-cut (BnC) method in  \cite{Yong2013}. The main idea of these two methods is to decompose the original problem into several more tractable subproblems and iteratively solve the subproblems until convergence. The condition for convergence is that the globally optimal solution can be obtained. Unfortunately, we consider the multichannel case, wherein each subproblem is non-convex and globally optimal solution cannot be obtained as explained in  \cite{cunhua2014}. Hence, the above two methods are not applicable. Instead, we adopt the exhaustive search method as the performance benchmark that is only simulated in a small network.}
 (denoted as `Exhaustive-search') is also shown, which evaluates every possible subset of UEs and chooses the feasible subset with the maximum number of UEs. Due to the exponential complexity associated with the exhaustive search, we only simulate a small network with $K=8$ and $I=12$. As expected, the number of admitted UEs for all algorithms decrease with the increase in the rate requirements. The optimal exhaustive search performs better than the `Joint-BUES-alg.', which comes at the cost of high computational complexity. However, the performance gap between these two algorithms is negligible when $R_{\rm{min}}$ is small (e.g., $R_{\rm{min}}<20\ {\rm{bit/s/Hz}}$). By jointly optimizing the beam direction and power allocation, the `Joint-BUES-alg.' outperforms the `MF-BUES-alg.'. However, the performance gain decreases when  $R_{\rm{min}}$ is large. The reason can be explained as follows. With the increase of $R_{\rm{min}}$, the number of admitted UEs decreases and these UEs are separated far away from each other. Hence, interference is not so significant and the channel matching beam direction approaches the optimal direction.

\subsubsection{Convergence behaviour of the NPC minimization algorithm}

Figure \ref{fig3} shows the convergence behaviour for the NPC minimization algorithm (i.e., Algorithm \ref{algorithmiter}), where `Joint-NPC-Alg.1' denotes the joint beam direction and power allocation optimization performed by Algorithm \ref{algorithmiter}, while `MF-NPC-Alg.1' denotes that the beam direction is fixed to be channel direction and the power optimization is carried out by Algorithm \ref{algorithmiter}. The top subplot shows the NPC trend, the middle and bottom subplots show the numbers of active RRHs and active links remained in each iteration, respectively. It is seen from this figure that all these values decrease rapidly and converge within twenty iterations. Both the convergence speed and the NPC performance for the considered two algorithms are very similar in this scenario. Moreover, for `Joint-NPC-Alg.1', the NPC, the number of active RRHs and active links decrease about $65\%$, $45\%$ and $94\%$, respectively, which confirm the effectiveness of the proposed algorithm in terms of power savings.

In the following, we will evaluate the effects of different system parameters on both the  NPC minimization algorithm (i.e.,Algorithm \ref{algorithmiter}) and UE selection algorithm (i.e., Algorithm \ref{bisecselece}). To compare the performance of the  NPC minimization algorithm, the performance of the conventional transmit power minimization is also considered, where all the RRHs in each UE's candidate set are assumed to be active. `Joint-Conven' and `MF-Conven' denote the conventional method when beam direction and power allocation are jointly optimized and beam direction is fixed at channel direction, respectively.

\subsubsection{Effects of the candidate size}

\begin{figure}
\begin{minipage}[t]{0.485\linewidth}
\centering
\includegraphics[width=2.8in]{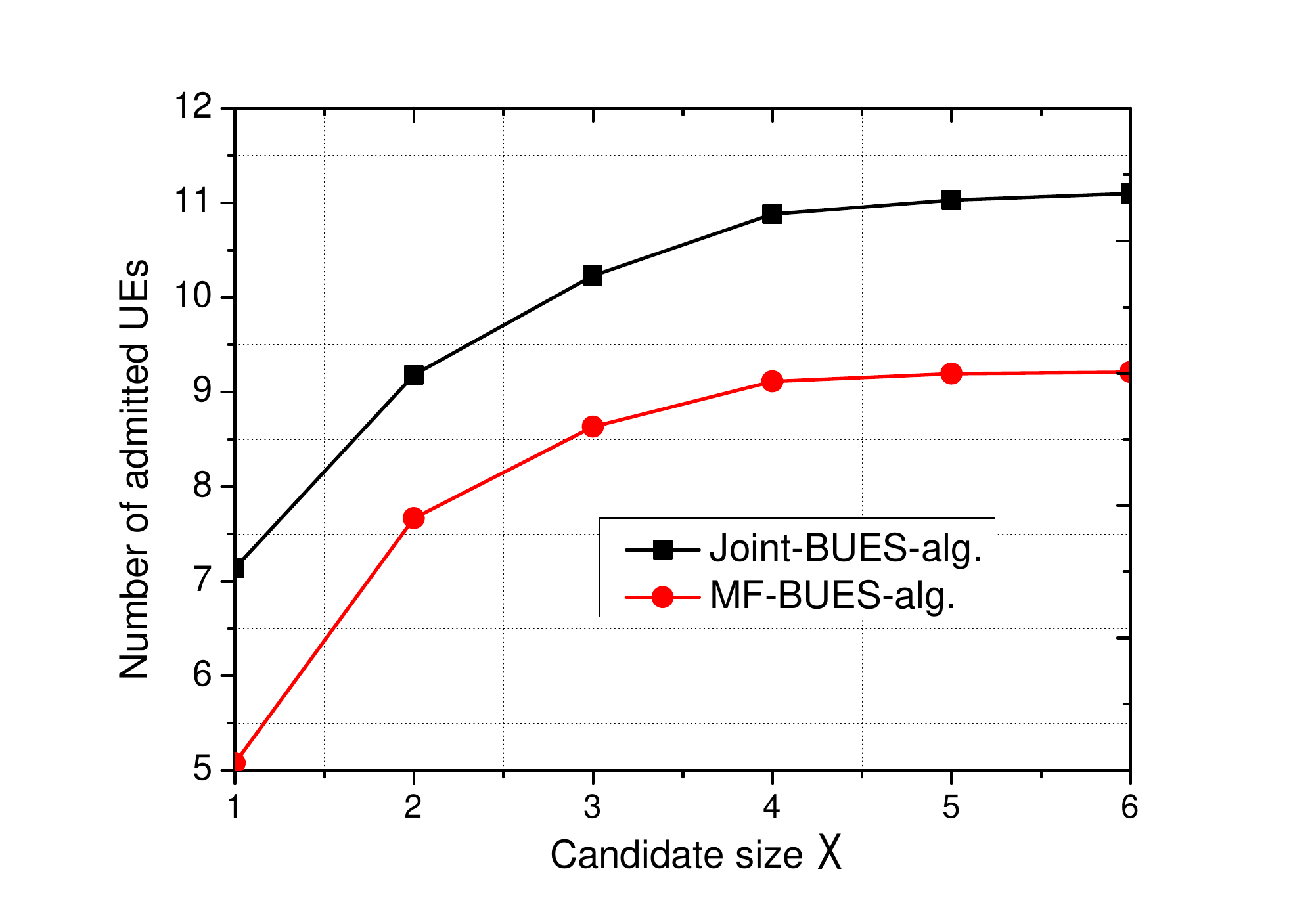}
\caption{Number of admitted UEs versus the candidate size $X$ with $Y=6$.}
\label{fig4}
\end{minipage}%
\hfill
\begin{minipage}[t]{0.485\linewidth}
\centering
\includegraphics[width=2.8in]{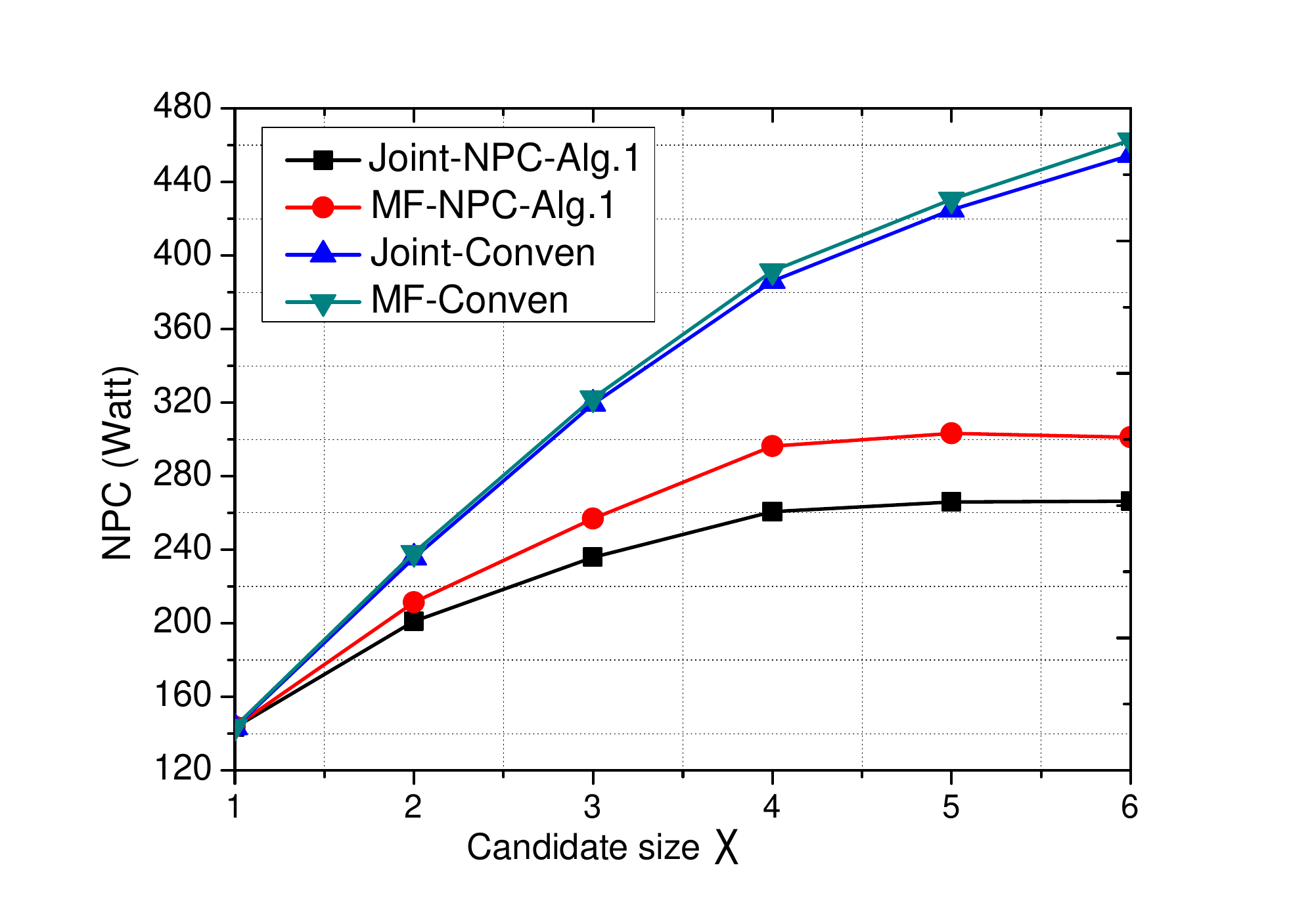}
\caption{NPC versus the candidate size $X$ for different algorithms with $Y=6$.}
\label{fig5}
\end{minipage}
\end{figure}

Figs.~\ref{fig4} and ~\ref{fig5} show the numbers of admitted UEs and NPC versus the candidate size $X$, respectively. The set of UEs that are admitted by the `MF-BUES-alg.' are set as the initialization point for the NPC minimization algorithms, which is the same in the following simulations. As expected, the larger candidate size leads to more admitted UEs due to the increasing network degrees of freedom. However, the number of admitted UEs achieved by `Joint-BUES-alg.' and `MF-BUES-alg.' become flat in the large candidate size regime, which is consistent with the conclusion in \cite{Penggenwcl2014}. This means that it is not necessary to consider the far away RRHs for each UE since they contribute less to their performance and the candidate size should not be larger than 4 to obtain a tradeoff between performance and implementation complexity. The similar trend holds for the `Joint-NPC-Alg.1' and `MF-NPC-Alg.1' in Figure~\ref{fig5}. However, in Figure~\ref{fig5}, the conventional transmit power minimization consumes much higher power than the proposed `Joint-NPC-Alg.1' and `MF-NPC-Alg.1', and the gap increases with the increase of candidate size. The reason is that with the increase of candidate size, more RRHs will be in the active mode, which requires large amount of circuit power consumption. It is surprising to see from Figure~\ref{fig5} that for the conventional method,  `MF-Conven' requires slightly higher power consumption than `Joint-Conven'. On the other hand, `MF-NPC-Alg.1' requires much higher power than `Joint-NPC-Alg.1'. These two facts confirm that joint beam direction and power allocation optimization is more important for RRH and link selection.

\subsubsection{Effects of the amount of CSI}

\begin{figure}
\begin{minipage}[t]{0.495\linewidth}
\centering
\includegraphics[width=2.8in]{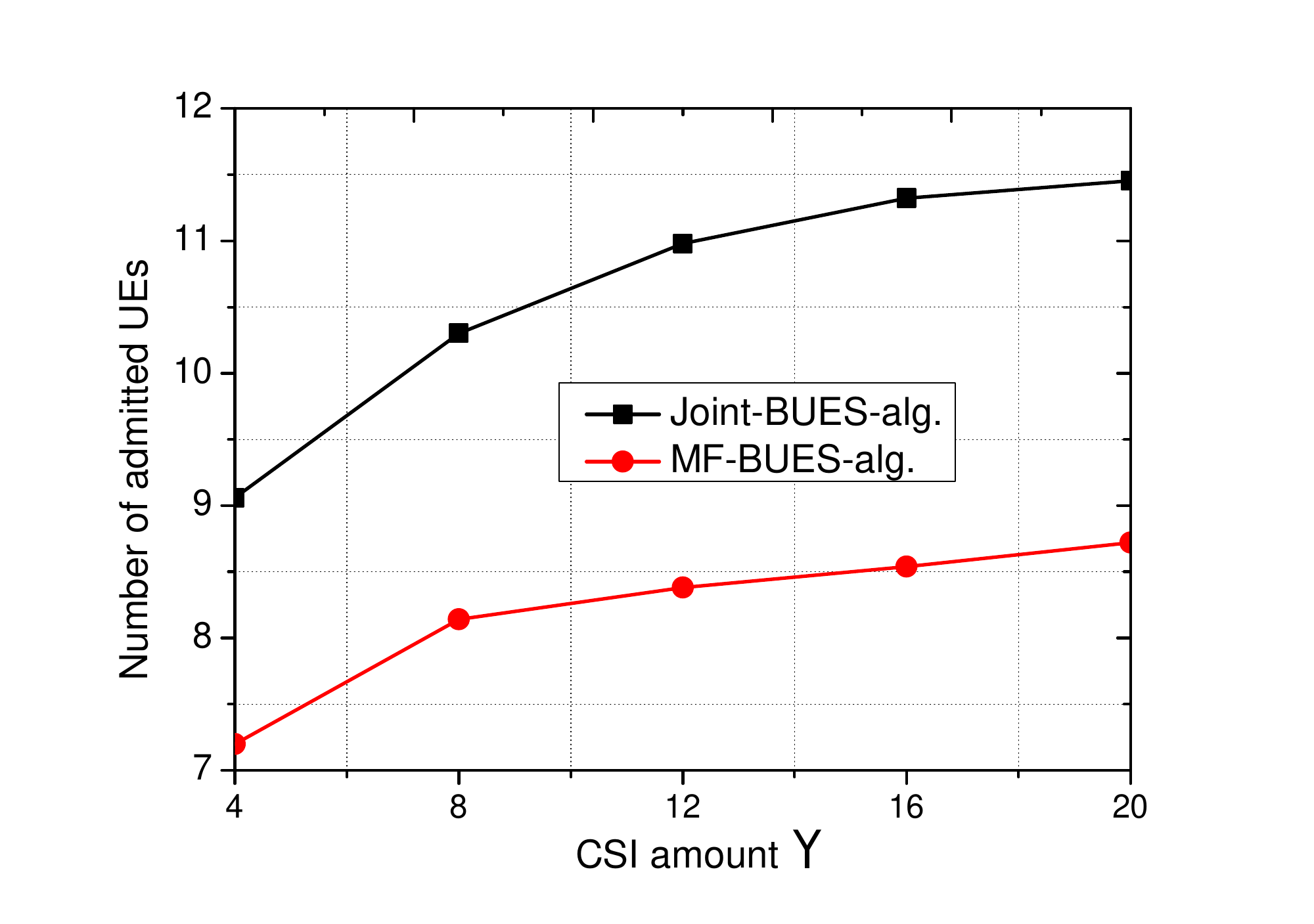}
\caption{Number of admitted UEs versus the CSI amount $Y$ with $X=3$. }
\label{fig6}
\end{minipage}%
\hfill
\begin{minipage}[t]{0.495\linewidth}
\centering
\includegraphics[width=2.8in]{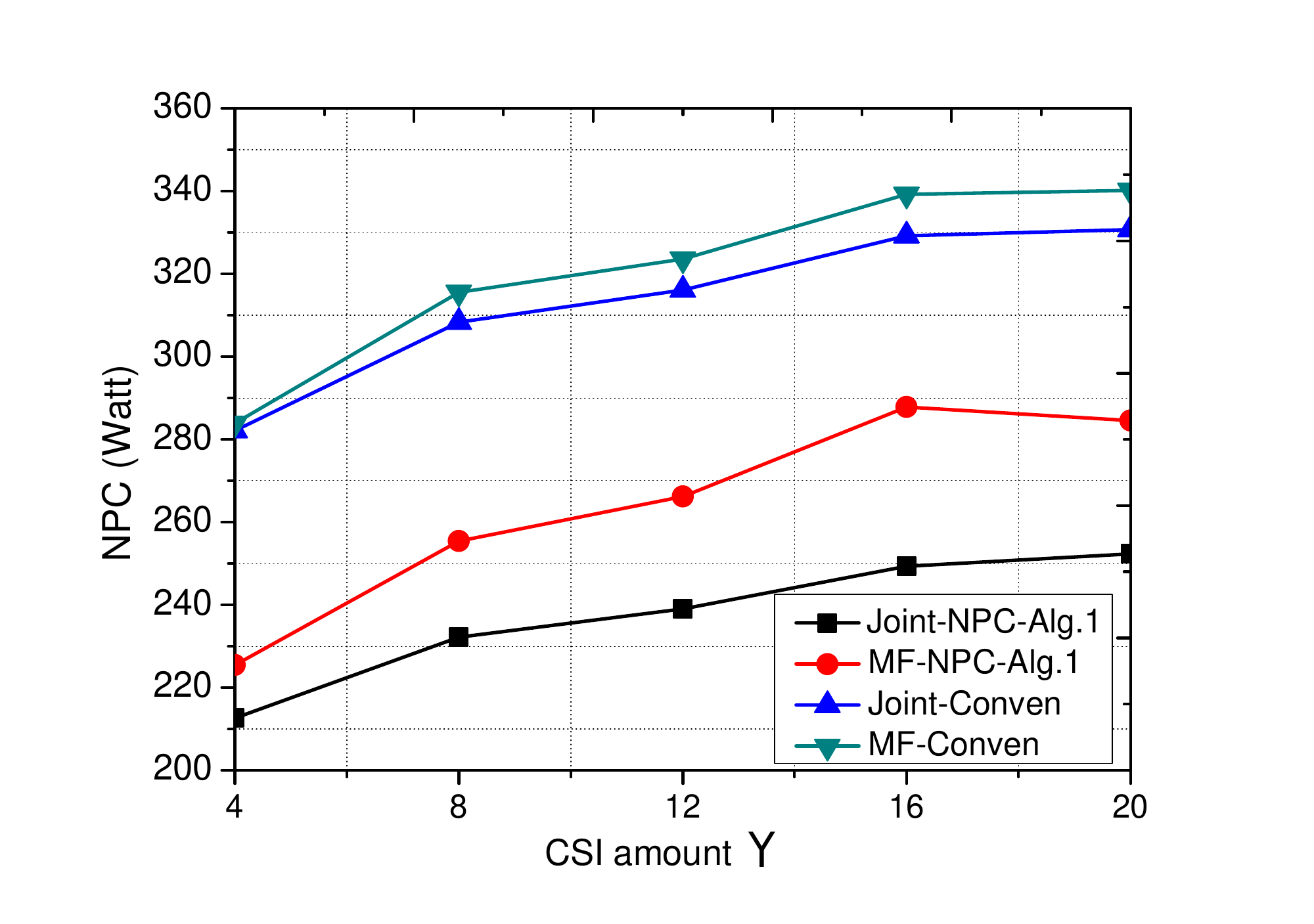}
\caption{NPC versus the CSI amount $Y$ for different algorithms with $X=3$.}
\label{fig7}
\end{minipage}
\end{figure}

Now, we investigate the effects of limited CSI on the performance of the proposed algorithms. Figs.~\ref{fig6} and \ref{fig7} illustrate the  number of admitted UEs and NPC versus the amount of CSI $Y$, respectively. Note that $Y$ denotes the number of (nearest) RRHs from which CSI is measured. As expected, the number of admitted UEs increase as the amount of CSI increases, since multi-user interference can be more accurately suppressed. From Figure \ref{fig6}, it is seen that the number of admitted UEs increases quickly when $Y<12$ and increases slowly in the high amount CSI regime. This result indicates that only a moderate amount of CSI is sufficient for the proposed algorithms to  achieve good performance, which can significantly reduce the channel estimation overhead. The corresponding NPC increases sightly with the amount of CSI due to more UEs are admitted. The proposed algorithms are again observed to perform much better than the conventional transmit power minimization method, highlighting the importance of joint optimization of transmit power, RRH and link selection.

\subsubsection{Effects of fronthaul capacity constraints}

\begin{figure}
\begin{minipage}[t]{0.495\linewidth}
\centering
\includegraphics[width=2.8in]{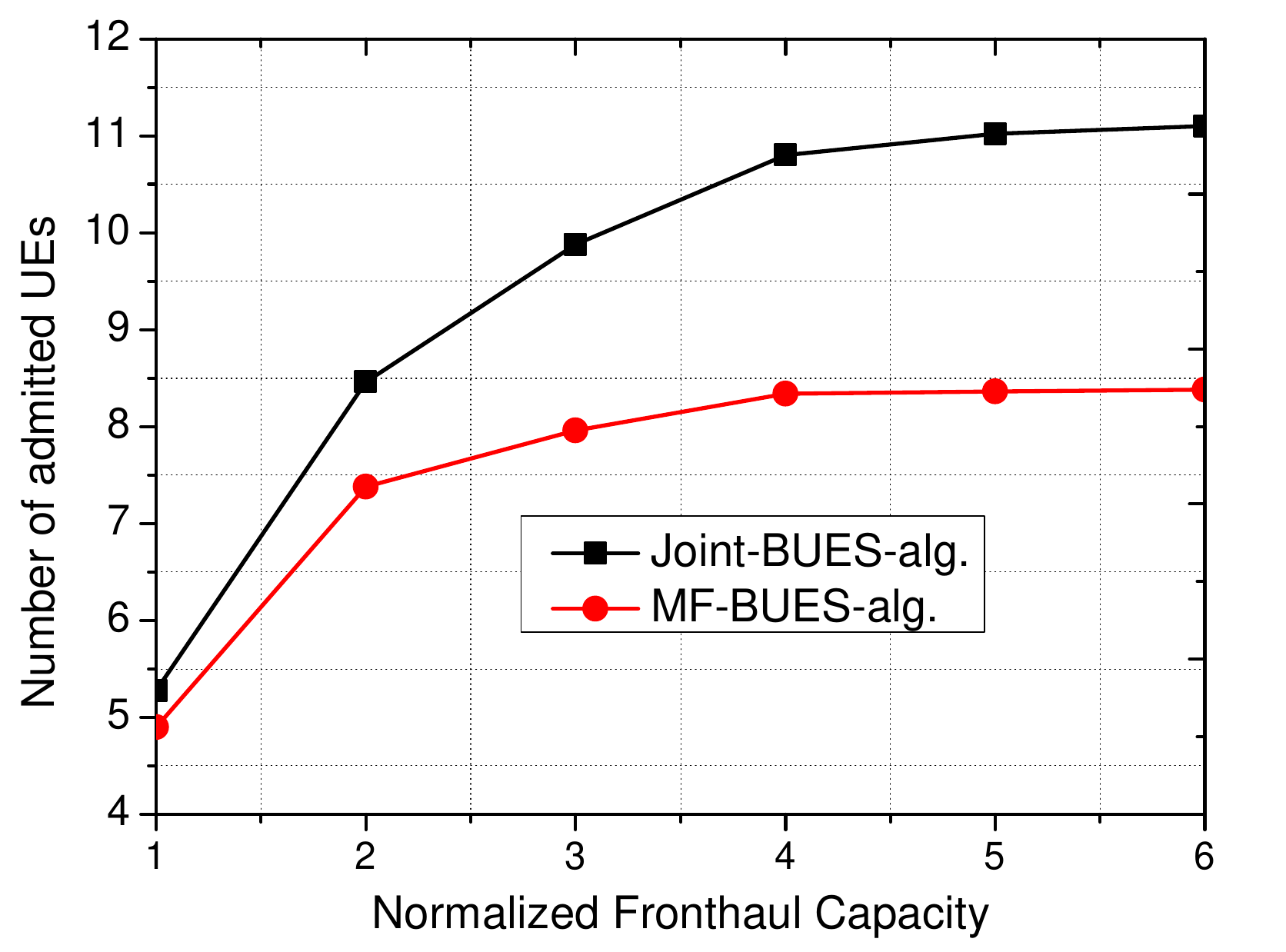}
\caption{Number of admitted UEs versus the normalized fronthaul capacity $\tilde C_{\rm{max}}$.}
\label{fig8}
\end{minipage}%
\hfill
\begin{minipage}[t]{0.495\linewidth}
\centering
\includegraphics[width=2.8in]{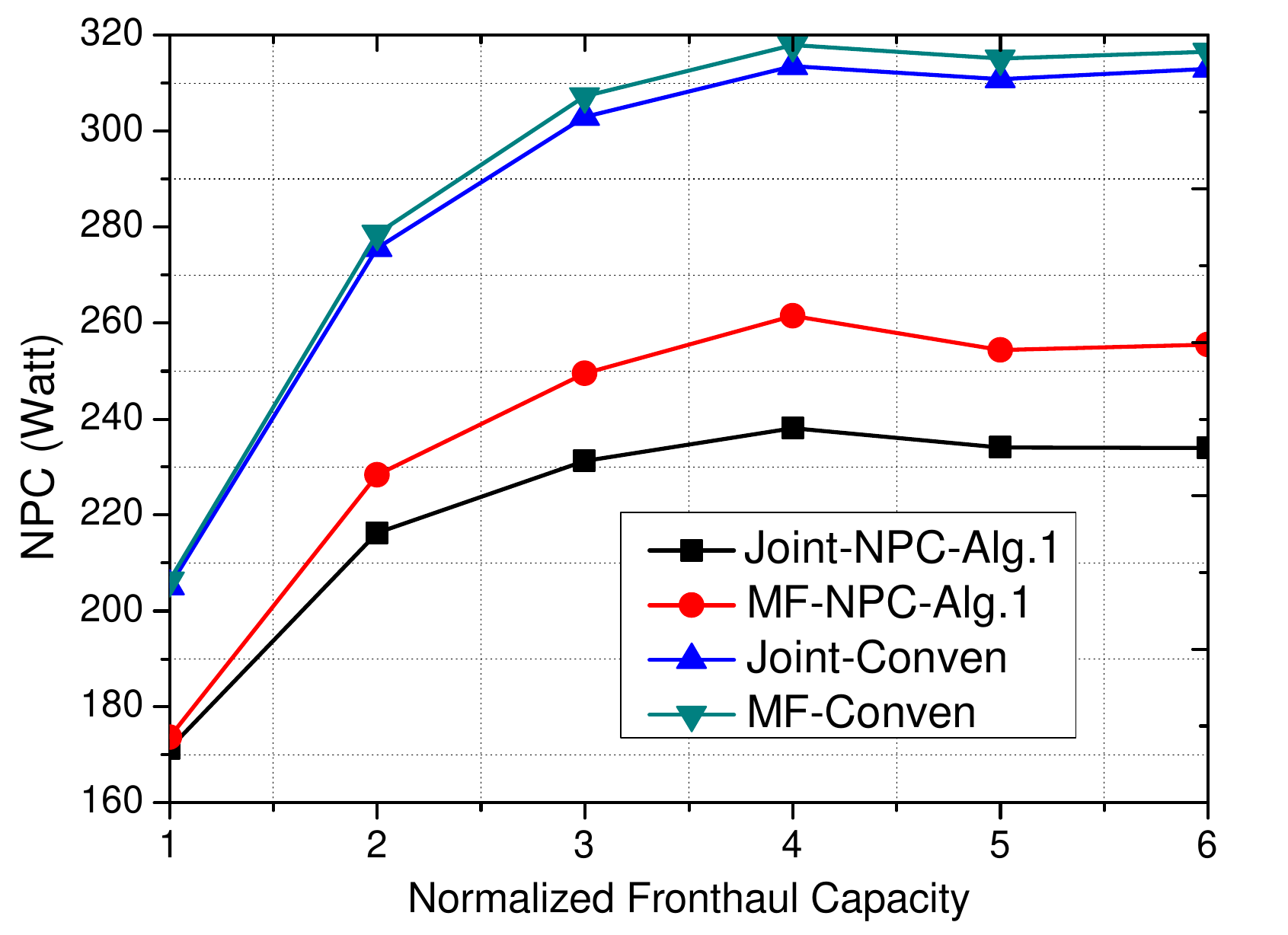}
\caption{NPC versus the normalized fronthaul capacity $\tilde C_{\rm{max}}$ for different algorithms.}
\label{fig9}
\end{minipage}
\end{figure}

Figs.~\ref{fig8} and \ref{fig9} show the number of admitted UEs and NPC versus the normalized fronthaul capacity $C_{\rm{max}}$, respectively. It is seen from Figure \ref{fig8} that the numbers of admitted UEs for both the  `Joint-BUES-alg.' and  `MF-BUES-alg.' increase with $C_{\rm{max}}$ initially due to the fact that  more UEs can be supported by each fronthaul link for large $C_{\rm{max}}$. However, the number of admitted UEs will be saturated in the large $C_{\rm{max}}$ regime. It is shown that  $C_{\rm{max}}=4$ is enough to achieve a large portion of the optimal performance, which indicates that the  fronthaul link capacity is not necessary to be very large and the wireless fronthaul link such as mmWave communication technologies may be applicable in dense C-RAN network. Figure \ref{fig8} also shows that `Joint-BUES-alg.' outperforms `MF-BUES-alg.' in terms of the number of admitted UEs and the performance gain increases with $C_{\rm{max}}$. From Figure \ref{fig9}, it can be seen that the NPC performances of `Joint-NPC-Alg.1' and `MF-NPC-Alg.1' increase with $C_{\rm{max}}$
when $C_{\rm{max}}<=4$, since more UEs are admitted in this regime as seen from Figure \ref{fig8}. However, when $C_{\rm{max}}>=4$, the NPC value experiences a slight decrease though the numbers of UEs are almost the same as seen in Figure \ref{fig8}. This is due to the fact that more flexibility in the fronthaul link can be exploited to reduce the numbers of active links and RRHs.

\section{Conclusions}\label{conclu}
This paper provided a complete framework to handle the challenges arising in the dense C-RAN. More specifically, the downlink beam-vectors, RRH selection and UE-RRH associations were jointly optimized to minimize the total NPC for dense C-RAN with incomplete CSI subject to fronthaul capacity constraints, UEs' QoS targets and per-RRH power constraints. We formulated this problem as an MINLP problem, which is NP-hard. In addition, the incomplete CSI makes the QoS constraints difficult to handle. We first replaced the exact expression of data rate with its lower bound. Then, we developed a low-complexity single-layer iterative algorithm to solve the NPC minimization problem based on the successive convex approximation technique and the equivalent relationship between data rate and MSE. Also, a low-complexity UE selection algorithm was proposed to guarantee the feasibility of the NPC problem. Simulation results showed that the proposed UE selection can achieve near-optimal performance compared to the optimal exhaustive UE search method. Moreover, the proposed single-layer iterative algorithm can achieve significant power savings in various setups. Simulation results also showed that only nearest four RRHs are sufficient to be the candidate set of each UE and limited CSI can  contribute large portion of performance gain from the full CSI case.

The future work lies in the joint optimization of cluster sizes and beam-vectors when taking into account the cost of computational complexity and channel training overhead. Also, it is worth studying how to extend the work to the scenario where each UE is equipped with multiple antennas.

\numberwithin{equation}{section}
\begin{appendices}
\section{Accurate closed-form expression of data rate for special case}\label{specialcase}
For the simplicity of notations, the SC index $n$ is omitted in the following derivations. The SINR for UE $k$ in (\ref{sinrre}) can be rewritten as
\begin{equation}\label{sinrepphk}
  {\gamma _k} = \frac{{{{\left| {{X_k}} \right|}^2}}}{{\sum\nolimits_{l \ne k,l \in {\cal U}} {{{\left| {{Y_{l,k}}} \right|}^2}}  + \sigma _k^2}},
\end{equation}
where ${X_k} = {{\bf{\bar h}}_{k,k}{\bf{\bar w}}_k}$ and $Y_{l,k}={{\bf{\bar h}}_{l,k}{\bf{\bar w}}_l}$. Note that ${\bf{\bar h}}_{k,k}$ is perfectly known and ${\bf{\bar w}}_l,\forall l$ are deterministic,  ${X_k}$ is a deterministic value and only $\{Y_{l,k},\forall l\in {\cal U}, l\neq k\}$ are random variables. According to the first two assumptions, all elements in ${\bf{\bar h}}_{l,k}$ are unknown and follow the circular symmetric complex Gaussian distribution. Specifically, the distribution of ${\bf{\bar h}}_{l,k}$ is given by ${\cal C}{\cal N}({\bf{0}},{\bf{R}}_{l,k})$, where ${\bf{R}}_{l,k}$ is a diagonal matrix. To obtain the expression of ${\bf{R}}_{l,k}$, we define the indices of ${{{\cal I}_l}}$ as ${{\cal I}_l} = \{ s_1^l, \cdots ,s_{|{{\cal I}_l}|}^l\} $. Then, based on the third assumption, ${\bf{R}}_{l,k}$ can be easily  calculated as ${{\bf{R}}_{l,k}} = {\rm{blkdiag}}\left( {{{\left| {{\alpha _{s_1^l,k}}} \right|}^2}{{\bf{I}}_{M \times M}}, \cdots ,{{\left| {{\alpha _{s_{|{{\cal I}_l}|}^l,k}}} \right|}^2}{{\bf{I}}_{M \times M}}} \right)$. Then, given beam-vector ${\bf{\bar w}}_l$,  $Y_{l,k}$ is a Gaussian random variable with zero mean and variance given by $\varpi_{l,k}={\bf{\bar w}}_l^H{{\bf{R}}_{l,k}}{{{\bf{\bar w}}}_l}$, i.e., $Y_{l,k}  \sim  {\cal {CN}}({\bf{0}},{\varpi _{l,k}})$. For convenience, denote ${Z_k} = \sum\nolimits_{l \ne k,l \in {\cal U}} {{{\left| {{Y_l}} \right|}^2}} $. Then, $Z_k$ follows a generalized chi-squared distribution, given by \cite{ye2016tradeoff}
\begin{equation}\label{jdoiew}
  f({z_k}) = \sum\nolimits_{l \ne k,l \in {\cal U}} {{T_{l,k}}{e^{ - {{{z_k}} \mathord{\left/
 {\vphantom {{{z_k}} {{\varpi _{l,k}}}}} \right.
 \kern-\nulldelimiterspace} {{\varpi _{l,k}}}}}}},
\end{equation}
where $T_{l,k}$ is given by
\[{T_{l,k}} = \frac{1}{{{\varpi _{l,k}}\prod\nolimits_{j \in {\cal U},j \ne l,k} {\left( {1 - \frac{{{\varpi _{j,k}}}}{{{\varpi _{l,k}}}}} \right)} }}.\]

Then, the data rate is derived as
 \begin{eqnarray}
&&\int_0^\infty  {{{\log }_2}} \left( {1 + \frac{{{{\left| {{X_k}} \right|}^2}}}{{{z_k} + \sigma _k^2}}} \right)f({z_k})d{z_k}\nonumber\\
 &&= \sum\limits_{l \ne k,l \in {\cal U}} {{T_{l,k}}} \int_0^\infty  {{{\log }_2}} \left( {1 + \frac{{{{\left| {{X_k}} \right|}^2}}}{{{z_k} + \sigma _k^2}}} \right){e^{ - \frac{{{z_k}}}{{{\varpi _{l,k}}}}}}d{z_k}\nonumber\\
&& = \sum\limits_{l \ne k,l \in {\cal U}} { - \frac{{{T_{l,k}}{\varpi _{l,k}}}}{{\ln 2}}} \int_0^\infty  {\left[ {\ln \left( {{z_k} + \sigma _k^2 + {{\left| {{X_k}} \right|}^2}} \right) - \ln \left( {{z_k} + \sigma _k^2} \right)} \right]} d{e^{ - \frac{{{z_k}}}{{{\varpi _{l,k}}}}}}\nonumber\\
&& = \sum\limits_{l \ne k,l \in {\cal U}} {\frac{{{T_{l,k}}{\varpi _{l,k}}}}{{\ln 2}}\!\!\left[ {\ln \left( {1 + \frac{{{{\left| {{X_k}} \right|}^2}}}{{\sigma _k^2}}} \right)\! +\! \int_0^\infty  {\frac{{{e^{ - \frac{{{z_k}}}{{{\varpi _{l,k}}}}}}}}{{{z_k} + \sigma _k^2 + {{\left| {{X_k}} \right|}^2}}}d{z_k}\! -\! \int_0^\infty  {\frac{{{e^{ - \frac{{{z_k}}}{{{\varpi _{l,k}}}}}}}}{{{z_k} + \sigma _k^2}}d{z_k}} } } \right]}\label{paritalintegra}\\
&& = \sum\limits_{l \ne k,l \in {\cal U}}\!\!\! {\frac{{{T_{l,k}}{\varpi _{l,k}}}}{{\ln 2}}\!\!\!\left[ {\ln\! \left(\! {1 + \frac{{{{\left| {{X_k}} \right|}^2}}}{{\sigma _k^2}}}\! \right)\! -\! {e^{\frac{{\sigma _k^2 + {{\left| {{X_k}} \right|}^2}}}{{{\varpi _{l,k}}}}}}{\rm{Ei}}\left( \!{ - \frac{{\sigma _k^2 \!+\! {{\left| {{X_k}} \right|}^2}}}{{{\varpi _{l,k}}}}}\! \right) + {e^{\frac{{\sigma _k^2}}{{{\varpi _{l,k}}}}}}{\rm{Ei}}\left(\! { - \frac{{\sigma _k^2}}{{{\varpi _{l,k}}}}} \!\right)} \right]}\label{finalexpre}
\end{eqnarray}
where ${\rm{Ei}}(x) =  - \int_{ - x}^\infty  {\left( {{e^{ - t}}/t} \right)dt} $ is an exponential integral function, (\ref{paritalintegra}) is obtained by using integration by parts, and (\ref{finalexpre}) is achieved by invoking [Eq. (3.352.4), \cite{jeffrey2007table}].

\section{Proof of Lemma 1}\label{prooflemma1}
 We prove that $\Psi_k^{(n)}\left( {{{\bf{w}}},q_k^{(n)},u_k^{(n)}} \right)$ is a lower bound of $\tilde r_k^{(n)}{({\bf{w}})}$ by showing that given ${{\bf{w}}}$, the maximum of $\Psi_k^{(n)}\left( {{{\bf{w}}},q_k^{(n)},u_k^{(n)}} \right)$  is equal to $\tilde r_k^{(n)}{({\bf{w}})}$.

Obviously, function $\Psi_k^{(n)}\left( {{{\bf{w}}},q_k^{(n)},u_k^{(n)}} \right)$ is respectively concave over $u_k^{(n)}$, $q_k^{(n)}$ and ${\bf{w}}$ when the other two are fixed. As a result, the optimal $u_k^{(n)}$ and  $q_k^{(n)}$ to achieve the maximum value of $\Psi_k^{(n)}\left( {{{\bf{w}}},q_k^{(n)},u_k^{(n)}} \right)$ are obtained by setting the first order of $\Psi_k^{(n)}\left( {{{\bf{w}}},q_k^{(n)},u_k^{(n)}} \right)$ to zero, which are given in (\ref{receU}) and (\ref{receW}), respectively.

By  inserting the expression of ${{u}}_k^{(n)\star}$ in (\ref{receU})  into (\ref{mse}), the expression of $\epsilon_k^{(n)}\left( {{\bf{u}}^{\star},{\bf{w}}} \right)$ can be obtained by (\ref{enk}). By substituting the optimal ${{u}}_k^{(n)\star}$ in (\ref{receU})  and ${{q}}_k^{(n)\star}$ in (\ref{receW})  into function $\Psi_k^{(n)}\left( {{{\bf{w}}},q_k^{(n)},u_k^{(n)}} \right)$, we have
\begin{eqnarray}
\Psi_k^{(n)}\left( {{{\bf{w}}},q_k^{(n)\star},u_k^{(n)\star}} \right)& =& {\log _2}e\ln {\left( {1 - \frac{{{{\left| {{\bf{\bar h}}_{k,k}^{(n)}{\bf{\bar w}}_k^{(n)}} \right|}^2}}}{{{{\left| {{\bf{\bar h}}_{k,k}^{(n)}{\bf{\bar w}}_k^{(n)}} \right|}^2} + \sum\limits_{l \in {\cal U},l \ne k} {{\bf{\bar w}}{{_k^{(n)}}^{\rm{H}}}{\bf{A}}_{l,k}^{(n)}} {\bf{\bar w}}_k^{(n)} + \sigma _k^2}}} \right)^{ - 1}}\\
 &=& {\log _2}\left( {1 + \frac{{{{\left| {{\bf{\bar h}}_{k,k}^{(n)}{\bf{\bar w}}_k^{(n)}} \right|}^2}}}{{\sum\limits_{l \in {\cal U},l \ne k} {{\bf{\bar w}}{{_k^{(n)}}^{\rm{H}}}{\bf{A}}_{l,k}^{(n)}} {\bf{\bar w}}_k^{(n)} + \sigma _k^2}}} \right)\\
 &{\rm{ = }}&\tilde r_k^{(n)}({\bf{w}}).
\end{eqnarray}
Hence, the proof is complete.

\section{Proof of Theorem 1}\label{prooftheorem1}
Before proving the theorem, we first construct the following auxiliary problem
\begin{subequations}\label{auxipro}
\begin{align}
{\cal P}_{X}:\ \mathop {\min }\limits_{{\bf{w}}} \quad
& {{\hat P}_{{\rm{tot,}}\theta }}\left( {\bf{w}} \right)
\\
\qquad\ \textrm{s.t.}\qquad\!\!\!\!
&{\rm{C3}},{\rm{C4}},{\rm{C8}}.
\end{align}
\end{subequations}
Note that Problem ${\cal P}_{X}$ has the same objective function as Problem ${\cal P}_{7}$. In the following, we show that Algorithm \ref{algorithmiter} actually solves Problem ${\cal P}_{X}$. In addition, the only difference between the original Problem ${\cal P}_{4}$  and Problem ${\cal P}_{X}$  is that the indicator function is replaced by the concave smooth function.

Since $\{{\bf{w}}{(0)}, \forall k\}$ is initialized by using the output from the UE selection algorithm, ${\bf{w}}{(0)}$ is a feasible solution of Problem ${\cal P}_{4}$. By using the fact that ${f_\theta }(x) < 1,\forall x$, we conclude that ${\bf{w}}{(0)}$ is also feasible for ${\cal P}_{X}$. Note that ${\bf{u}}{(0)}$ and ${\bf{q}}{(0)}$ are calculated by using (\ref{receU}) and (\ref{receW}) with ${\bf{w}}{(0)}$. Then by using Lemma 1, ${\bf{w}}{(0)}$ is a feasible solution of Problem ${\cal P}_{7}$ with fixed ${\bf{u}}{(0)}$ and ${\bf{q}}{(0)}$. It is easy to check that ${\bf{w}}{(0)}$ is also a feasible solution of Problem ${\cal P}_{8}$ with $\{\beta _i{(0)}, \chi _{i,k}{(0)},  {\bf{G}}_k{(0)},  {\tau _{i,k}}(0), {{\tilde C}_{i,\max }}(0), \forall i,k \}$. Now, we consider step 2 of the first iteration (i.e., $t=1$) of Algorithm \ref{algorithmiter}. Since ${\bf{w}}{(1)}$ is the optimal solution of Problem ${\cal P}_{8}$, we have
\begin{equation}\label{decreasi}
 \begin{array}{l}
 \sum\limits_{n \in {\cal N}} {\sum\limits_{k \in {\cal U}} {{\bf{\bar w}}_k^{(n){\rm{H}}}(1){{\bf{G}}_k}(0){\bf{\bar w}}_k^{(n)}(1)} }
 \le  \sum\limits_{n \in {\cal N}} {\sum\limits_{k \in {\cal U}} {{\bf{\bar w}}_k^{(n){\rm{H}}}(0){{\bf{G}}_k}(0){\bf{\bar w}}_k^{(n)}(0)} }.
\end{array}
\end{equation}

For the simplicity of representation, denote ${\psi _i}(t) = P_i^{{\rm{tr}}}({\bf{w}}(t))$ and ${\xi _{i,k}}(t) = P_{i,k}^{{\rm{tr}}}({\bf{w}}(t))$. Then,  we have
\begin{align}
&{\rm{Obj}}({{\bf{w}}{(1)}})\nonumber\\
& = \sum\limits_{i \in {{\cal I}}} {\left( {{{{\eta _i}}}{\psi _i}(1)  + {f_\theta }\left( {\psi _i}(1)  \right)P_i^{\rm{c}} + {\rho _i}\sum\limits_{k \in {{\cal U}_i}} {{f_\theta }\left( {\xi_{i,k}{(1)}} \right){R_{k,\min }}} } \right)} \nonumber\\
&\mathop  \le \limits^{(a)}  \sum\limits_{i \in {\cal I}} {\left( {{\eta _i}{\psi _i}(1) + {f_\theta }\left( {{\psi _i}(0)} \right)P_i^{\rm{c}} + {\beta _i}(0)P_i^{\rm{c}}\left( {{\psi _i}(1) - {\psi _i}(0)} \right)} \right)}+  \nonumber\\
&\quad\sum\limits_{i \in {\cal I}} {\sum\limits_{k \in {{\cal U}_i}} {{\rho _i}{R_{k,\min }}\left( {{f_\theta }\left( {{\xi _{i,k}}(0)} \right) + { \chi _{i,k}}(0)\left( {{\xi _{i,k}}(1) - {\xi _{i,k}}(0)} \right)} \right)} } \nonumber\\
&\mathop  \le \limits^{(b)} \sum\limits_{i \in {\cal I}} {\left( {{\eta _i}{\psi _i}(0) + {f_\theta }\left( {{\psi _i}(0)} \right)P_i^{\rm{c}} + {\beta _i}(0)P_i^{\rm{c}}\left( {{\psi _i}(0) - {\psi _i}(0)} \right)} \right)}+ \nonumber\\
&\quad \sum\limits_{i \in {\cal I}} {\sum\limits_{k \in {{\cal U}_i}} {{\rho _i}{R_{k,\min }}\left( {{f_\theta }\left( {{\xi _{i,k}}(0)} \right) + { \chi _{i,k}}(0)\left( {{\xi _{i,k}}(0) - {\xi _{i,k}}(0)} \right)} \right)} } \nonumber\\
& = \sum\limits_{i \in {{\cal I}}} {\left( {{{{\eta _i}}}{\psi _i}(0)  + {f_\theta }\left( {\psi _i}(0)  \right)P_i^{\rm{c}} + {\rho _i}\sum\limits_{k \in {{\cal U}_i}} {{f_\theta }\left( {\xi_{i,k}{(0)}} \right){R_{k,\min }}} } \right)} \nonumber \\
& = {\rm{Obj}}({{\bf{w}}{(0)}})\nonumber
\end{align}
where ${\rm{Obj}}({{\bf{w}}{(t)}})$ denotes the objective value of Problem ${\cal P}_{7}$ or ${\cal P}_{X}$, (a) follows by using (\ref{first}) and (\ref{second}), (b) follows due to (\ref{decreasi}).

 Next, we show that ${\bf{w}}{(1)}$ is also a feasible solution of Problem ${\cal P}_{X}$. Obviously, ${\bf{w}}{(1)}$ satisfies C3, i.e., the power constraints. We only need to prove that ${\bf{w}}{(1)}$ satisfies C4 and C8.

 Since ${\bf{w}}{(1)}$ is the optimal solution of Problem ${\cal P}_8$, we have
 \begin{equation}\label{hhdss}
   \sum\nolimits_{n \in {\cal N}} {\Psi _k^{(n)}\left( {{{\bf{w}}}(1),q_k^{(n)}(0),u_k^{(n)}(0)} \right)}  \ge {R_{k,\min }}.
 \end{equation}
 In step 4 of the first iteration of Algorithm \ref{algorithmiter}, ${\bf{u}}{(1)}$ and ${\bf{q}}{(1)}$ are updated by using (\ref{receU}) and (\ref{receW}) with  ${\bf{w}}{(1)}$. Then according to Lemma 1, we have
 \begin{align}
 \sum\limits_{n \in {\cal N}} {\tilde r_k^{(n)}} ({\bf{w}}(1)) &= \sum\limits_{n \in {\cal N}} {\Psi _k^{(n)}\left( {{\bf{w}}(1),q_k^{(n)}(1),u_k^{(n)}(1)} \right)} \\
 &\ge \sum\limits_{n \in {\cal N}} {\Psi _k^{(n)}\left( {{\bf{w}}(1),q_k^{(n)}(0),u_k^{(n)}(0)} \right)} \\
 & \ge {R_{k,\min }}.
 \end{align}
 Hence, C4 is satisfied.

In addition, we have
\begin{align}
{C_{i,\max }} &\ge\sum\limits_{k \in {{\cal U}_i}} {\left( {{f_\theta }\left( {{\xi _{i,k}}(0)} \right) + { \chi _{i,k}}(0)\left( {{\xi _{i,k}}(1) - {\xi _{i,k}}(0)} \right)} \right){R_{k,\min }}} \label{equfirstone} \\
 &\ge \sum\limits_{k \in {{{\cal U}}_i}} {{f_\theta }\left( {{\xi_{i,k}{(1)}}} \right){R_{k,\min }}}, \label{equasectwo}
\end{align}
where (\ref{equfirstone}) follows since $ {\bf{w}}{(1)} $ is the solution of Problem ${\cal P}_8$  given ${\bf{u}}{(0)}$ and ${\bf{q}}{(0)}$,  and (\ref{equasectwo}) follows by using (\ref{second}). Hence, C8 is satisfied.

As a result, we can conclude that ${\bf{w}}{(1)}$ is also feasible for Problem ${\cal P}_{X}$. By using the similar method, we can obtain
\begin{equation}\label{links}
 {\rm{Obj}}({{\bf{w}}{(0)}}) \ge {\rm{Obj}}({{\bf{w}}{(1)}}) \ge {\rm{Obj}}({{\bf{w}}{(2)}}) \ge  \cdots .
\end{equation}
Obviously, the objective value  of Problem ${\cal P}_7$ (also ${\cal P}_{X}$) is lower bounded by zero. Hence, Algorithm \ref{algorithmiter} is guaranteed to converge in objective values.

Next, we prove the second part of the theorem: Given the feasible input ${\bf{w}}(0)$, the solution obtained from Algorithm \ref{algorithmiter} will converge to a unique point. Obviously, when $\bf{w}$ is given, ${\bf{u}}$ and ${\bf{q}}$  can be uniquely determined by using (\ref{receU}) and (\ref{receW}), respectively. Since $\{{\bf{G}}_k, \forall k\}$ are positive definite matrices, the objective function in Problem  ${\cal P}_8$ is a strictly convex function  of $\bf{w}$. Furthermore, it can be easily proved that the constraints in Problem  ${\cal P}_8$ are convex. As a result, Problem  ${\cal P}_8$ is a strictly convex optimization problem.
According to [Page 137 in \cite{boyd2004convex}], the globally optimal solution is unique. Then, by iteratively  updating step 2 to step 4 in Algorithm \ref{algorithmiter},  the algorithm will converge to a unique solution. We emphasize that since Problem ${\cal P}_{X}$ is a non-convex optimization problem, it may have multiple locally optimal solutions and its converged unique solution depends on the initial point. However, once the initial point is given, Algorithm \ref{algorithmiter} will converge to a unique solution.

\end{appendices}

\
\






\vspace{-1cm}
\bibliographystyle{IEEEtran}
\bibliography{myre}


\end{document}